\documentclass[12pt, draftclsnofoot, onecolumn]{IEEEtran}
\usepackage{amsfonts}
\usepackage{amsmath}
\usepackage{amsthm}
\usepackage{amssymb}
\usepackage{amsfonts}
\usepackage{amsmath}
\usepackage{amsthm}
\usepackage{graphicx}
\usepackage{epstopdf}
\usepackage{subcaption}
\usepackage[linesnumbered,ruled,lined]{algorithm2e}

\usepackage{color}
\usepackage{cite}
\DeclareMathOperator{\Trn}{\mathsf{T}}
\DeclareMathOperator{\Hrm}{\mathsf{H}}
\allowdisplaybreaks
\newtheorem{theo}{Theorem}
\newtheorem{lemma}{Lemma}
\newtheorem{definition}{Definition}
\newtheorem{proposition}{Proposition}
\newtheorem*{remark}{Remark}
\begin{document}
\title{Rate-Splitting Multiple Access for Overloaded Cellular Internet of Things}

\author{\IEEEauthorblockN{Yijie~Mao, \IEEEmembership{Member, IEEE}, Enrico~Piovano, and~Bruno~Clerckx, \IEEEmembership{Senior Member, IEEE}}
    
\thanks{
    	This work has been partially supported by the U.K. Engineering and Physical
    	Sciences Research Council (EPSRC) under grant EP/N015312/1, EP/R511547/1.
    	A preliminary version of this paper was presented at the  50th Asilomar Conference on Signals, Systems and Computers (ACSSC) \cite{enrico2016bruno}.}}

\maketitle
\begin{abstract}
	In the near future, it is envisioned that cellular networks will have to cope with extensive Internet of Things (IoT) devices. Therefore, a required feature of cellular IoT will be the capability to serve simultaneously a large number of devices with heterogeneous demands and qualities of Channel State Information at the Transmitter (CSIT). In this paper, we focus on an overloaded Multiple-Input Single-Output (MISO) Broadcast Channel (BC) with two groups of CSIT qualities, namely one group of users (representative of high-end devices) for which the transmitter has partial knowledge of the CSI, the other group of users (representative of IoT devices) for which the transmitter only has knowledge of the statistical CSI.  We introduce Rate-Splitting Multiple Access (RSMA), a new multiple access based on multi-antenna Rate-Splitting (RS) for cellular IoT.  Two strategies are proposed, namely, Time Partitioning--RSMA (TP--RSMA)  and Power Partitioning--RSMA (PP--RSMA). The former  independently serves the two groups of users over orthogonal time slots while the latter  jointly serves the two groups of users within the same time slot in a non-orthogonal manner. We first show at high Signal-to-Noise Ratio (SNR) that PP--RSMA achieves the optimum Degrees-of-Freedom (DoF) in an overloaded MISO BC with heterogeneous CSIT qualities and then  show at finite SNR that, by marrying the benefits of PP and RSMA,  PP--RSMA  achieves explicit  sum rate gain over TP--RSMA and all baseline schemes. Furthermore, PP--RSMA is robust to CSIT inaccuracy and flexible to cope with Quality of Service (QoS) rate constraints of all users.  The DoF and rate analysis helps us draw the conclusion that PP--RSMA  is a powerful framework for  cellular IoT with a large number of devices.
	
\end{abstract}
\section{Introduction} \label{sec: introduction}
%
 The upsurge of urban population and the inevitable trend of urbanization all over the world result in an urgent demand on sustainable city transition, which has driven the idea of ``smart city" \cite{IoT2014Jin}. As an indispensable platform for cities to be smarter, Internet of Things (IoT)  provides advanced solutions in the realms of smart grids, smart homes, smart transportations, etc, with the assistance of a large number of IoT devices. The expeditious development of IoT in different fields introduces multifarious criteria for IoT network design such as   massive connectivity, security, trustworthy, ultra-low latency, throughput, ultra-reliability, etc, which in fact imposes dramatic pressure on wireless access networks \cite{5GIoT2018survey, chen2019massive}. To explore  solutions to meet all those criteria, the concept of \textit{cellular IoT} is introduced by the 3rd Generation Partnership Project (3GPP) for Long Term Evolution (LTE) \cite{3gpp45820}. 

Though 5G cellular IoT is still in its infancy, a key enabling technology that has been widely recognized is the advanced spectrum sharing and interference management via Multi-User Multiple-Input Multiple-Output (MU-MIMO) \cite{5GIoT2018survey}. By leveraging  multiple antennas at the transmitter, the spatial resources are exploited which open the door to a well-established multiple access technique in current wireless communication networks, namely,  Space Division Multiple Access (SDMA). By superposing users in the same time-frequency resources and separating them through spatial domain, SDMA is capable of boosting the system  spectral efficiency and  achieving the optimal Degree-of-Freedom (DoF) under perfect Channel State Information at the Transmitter (CSIT) \cite{RSintro16bruno}.  Though appealing in its concept, the following drawbacks of SDMA limit its application to cellular IoT. 
The implementation of SDMA  using  Multi-User Linear Precoding (MU--LP) is  only suited to an underloaded regime and requires user channel to be (semi-)orthogonal with accurate CSIT. The performance of SDMA drops dramatically when the network  becomes overloaded, or when the user channels lose orthogonality, or when the CSIT turns to be imperfect. The achievable DoF of SDMA under imperfect CSIT is in fact suboptimal since residual interference is introduced by the distorted interference nulling. To tackle those issues, additional user scheduling is required, which would increase latency and impede a larger number of IoT connectivity if the transmitter has a limited number of transmit antennas \cite{mao2017rate}. 

An alternative multiple access scheme that has been studied in cellular IoT is  power-domain Non-Orthogonal Multiple Access (NOMA)\footnote{In the sequel, ``power-domain NOMA" is referred by NOMA for simplicity.  } \cite{Shirvan2017MassiveNOMA,Chen2018FullyNOMA}. The original NOMA which relies on Superposition Coding (SC) at the transmitter and Successive Interference Cancellation (SIC) at the receivers (also known as SC--SIC)  has been recognized as the capacity-achieving strategy of Single-Input Single-Output (SISO) Broadcast Channel (BC) \cite{Tcover1972}.
A simple application of SC--SIC to Multiple-Input Single-Output (MISO) BC, however, would hamper the DoF to unity since one receiver has to decode  data streams of all users. Moreover, the receiver complexity is extremely high as the number of SIC deployed at each user scales with the number of users in the system.  To compensate the disadvantages of SC--SIC, efforts have been made to another strategy based on SC--SIC where users are scheduled into different user groups with users in the same group being served by SC--SIC  while users across groups being served by SDMA. Although the DoF increases and the receiver complexity decreases by using such strategy, additional limitations are introduced.  Users within each group are required to have aligned channels and users across  groups should have orthogonal channels. It is still sensitive to CSIT inaccuracy since the inter-group interference is managed by SDMA. Meanwhile, the scheduling complexity  dramatically increases due to the issue of  user grouping together with decoding order optimization. Now that users within each group are served by SC--SIC, SIC is still required at each IoT device. As most of the IoT devices are characterized by small-size, low-power nodes with simple function units \cite{chen2019massive}, deploying SIC(s) at each IoT device is actually impractical. 

One major challenge of MIMO networks that has been widely discussed is  the CSIT acquisition \cite{hassibi2003TDD, DLove2009FDD}. Imperfect channel estimation at the transmitter aggravates SDMA and NOMA to support extensive devices in cellular IoT. 
However, due to the uplink channel estimation error caused by quantized feedback in Frequency Division Duplex (FDD) systems or feedback delay in Time Division Duplex (TDD), CSIT is inevitably imperfect. In cellular IoT, such phenomenon would be more severe since the IoT devices  have limited functionalities. Moreover, the channels of highly-mobile IoT devices (e.g., Unmanned Aerial Vehicle (UAV) and vehicular equipments)  change rapidly. Even instantaneous imperfect CSIT cannot be reached and only statistical CSI is known at the transmitter \cite{chen2019massive}. 

In this work, motivated by all the aforementioned challenges, we design two novel transmission techniques  based on  \textit{Rate-Splitting Multiple Access (RSMA)} for cellular IoT. RSMA has been  recognized as a powerful and generalized physical-layer transmission framework for downlink multi-antenna BC that encompasses (and outperforms) SDMA and NOMA (including SC--SIC and SC--SIC per group) as sub-schemes \cite{mao2017rate,mao2018networkmimo,mao2018EE,bruno2019wcl,mao2019TCOM,mao2019beyondDPC}.  By splitting the messages of users into private and common parts, jointly encoding the common parts into  common streams to be decoded by multiple users and independently encoding the private parts into the private streams to be decoded by the corresponding users only, Rate-Splitting (RS), as the building block of RSMA, enables to partially decode the interference and partially treat the remaining  interference as noise. This contrasts sharply with SDMA that relies on fully treating interference as noise and NOMA that relies on fully decoding the interference \cite{mao2017rate}. 
Additionally, what is remarkable with RS-based strategies is their robustness to imperfect CSIT in multi-antenna BC. While the capacity region of MIMO BC with perfect CSIT is known and achieved with Dirty Paper Coding (DPC), that with imperfect CSIT remains unknown. However, RS has been discovered to be DoF optimal \cite{RS2016hamdi,enrico2017bruno,hamdi2019spawc} and outperforms DPC in underloaded multi-antenna BC with imperfect CSIT \cite{mao2019beyondDPC}.
 The optimality of RS in the DoF sense motivates recent studies of precoder design  in multi-antenna BCs at finite Signal-to-Noise Ratio (SNR) where RS has been shown to achieve higher spectral and energy efficiencies than SDMA, NOMA, Orthogonal Multiple Access (OMA),  and multicast in any network loads, CSIT inaccuracy and user deployments with a diversity of channel strength disparities and channel directions in MISO BC \cite{mao2017rate,RS2016hamdi}, non-orthogonal unicast and multicast transmission \cite{mao2019TCOM}, Cloud Radio Access Networks (C-RAN) \cite{Ahmad2018SPAWC,cran2019wcl}, multi-group multicast \cite{hamdi2017bruno}, massive MIMO  \cite{Minbo2016MassiveMIMO}, millimeter-wave systems \cite{minbo2017mmWave}, Simultaneous Wireless Information and Power Transfer (SWIPT) \cite{mao2019swipt},  cooperative RS in MISO BC with user relaying \cite{jian2019crs,mao2019maxmin}, etc. 
However, most of the above works focus on an underloaded regime where the number of messages is less than or equal to the number of transmit antennas, and with equal CSIT qualities among users (either  with perfect CSIT only or with imperfect CSIT only).  As the networks of cellular IoT is envisioned to serve simultaneously a large number of devices with heterogeneous CSIT qualities and demands,  it is expected that many networks will
operate in overloaded regimes, roughly described as scenarios where the number of messages exceeds the number of transmitting antennas. One fundamental example is captured by SISO BC widely studied in the literature.

\textit{Contributions:}
In this work, we focus on an \textit{overloaded} MISO BC with two groups of CSIT qualities, where there is a group of high-end users for which the transmitter obtains instantaneous imperfect CSI in each fading state and users are powerful enough to perform SIC and another group of users are simple-functioned IoT devices for which the transmitter only knows statistical CSI and the users cannot perform SIC. 
Based on the considered overloaded MISO BC with heterogeneous CSIT, we make the following four major contributions:

\begin{enumerate}
\item We propose two transmission frameworks in the realm of RSMA, namely, Time Partitioning (TP)--RSMA and Power Partitioning (PP)--RSMA, where TP--RSMA independently serves the two groups of users over orthogonal time slots while  PP--RSMA non-orthogonally serves the two groups of users within the same time slots. Both  are studied in  finite and high SNR regimes.
\item In the high SNR regime,  the achievable DoF of the proposed RSMA approaches is characterized. We show that  PP--RSMA  achieves strict DoF gains over TP--RSMA. Most significantly,  PP--RSMA  is shown to achieve the optimum DoF region in the considered overloaded MISO BC with heterogeneous CSIT, which contrasts with SDMA  and NOMA strategies  achieving suboptimal DoF only.
\item In the finite SNR regime,  we study the achievable sum rate  of the considered overloaded MISO BC with heterogeneous CSIT. With the objective of maximizing the   Ergodic Sum Rate (ESR) of the high-end users over a long sequence of fading states subject to  the Quality of Service (QoS) rate constraints of all users, we optimize the precoders and demonstrate the proposed PP--RSMA achieves explicit sum rate gain over TP--RSMA and other  baseline schemes, i.e., PP/TP--SDMA. The anticipated gain of PP--RSMA over TP--RSMA at high SNR is  obtained at finite SNR as well. 
\item  Furthermore, we study the effect of network load, CSIT inaccuracy,  QoS rate requirements of all users on the system performance. We show through  numerical results that the ESR gain of PP-based strategies  over TP-based strategies increases dramatically with the number of IoT users or  the QoS rate requirement of those users. RSMA-based strategies achieve higher ESR gain over SDMA-based strategies as the QoS rate requirement of high-end users increases. By marrying the benefits of PP and RSMA, PP--RSMA becomes a powerful approach for cellular IoT as it is less sensitive to CSIT inaccuracy and achieves a higher spectral efficiency. 
\end{enumerate}

\textit{Organizations:} The rest of the paper is organized as follows. The system
model is described in Section \ref{sec: system_model}. The proposed RSMA approaches are specified in Section \ref{sec: RSMA} followed by the  DoF analysis in Section \ref{sec: DoF} and problem formulation in Section \ref{sec: problem formulation}.  The optimization framework is discussed in Section  \ref{sec: optimization framework}. Section \ref{sec: numerical results} provides the numerical results and  Section \ref{sec: conclusion} concludes the paper.

\textit{Notations:} We use bold upper (lower) letters to denote matrices (column vectors).  $\mathbf{I}$ denotes the identity matrix. 
{$(\cdot)^{\Trn}$,   $(\cdot)^{\Hrm}$},  $\|\cdot\|$, $\mathbb{E}\{\cdot\}$,  $\mathrm{tr}(\cdot)$   represent the transpose, Hermitian,  Euclidean norm, expectation and trace  operators, respectively.  $\mathcal{CN}(0,\sigma^2 )$ denotes the Circularly Symmetric Complex Gaussian (CSCG) distribution with zero mean and variance $\sigma^2$.  
\section{System Model } \label{sec: system_model}
We consider a wireless network where a transmitter with $M$ antennas (i.e. Base Station, BS) serves $K$ single-antenna receivers 
(i.e. users). The set of user indices is given by $\mathcal{K}=\{1,\ldots,K\}$.
We focus on the regime $K>M$, and refer to the underlying channel  as the overloaded MISO BC.
Communication occurs over $T$ uses of the wireless channel, where channel uses are indexed by $\mathcal{T}=\{1,\ldots,T\}$,
and the input-output relationship at the $t$-th channel use (or time instance) is described by
\begin{equation}
	\label{eq: receivedSignal}
	y_k(t)=\mathbf{h}_k^{\Hrm}(t)\mathbf{x}(t)+n_k(t), \ \forall k\in \mathcal{K},  t\in\mathcal{T}.
\end{equation}
In the above,  $\mathbf{x}(t)\in\mathbb{C}^{M\times 1}$ is the transmit signal, subject to a per-codeword average power constraint given by $\frac{1}{T} \sum_{t = 1}^{T }\|\mathbf{x}(t)\|^2 \leq P$; $y_k(t)$ is the signal received by user $k$; 
$n_k(t)\sim \mathcal{CN}(0,1)$ is the corresponding  Additive White Gaussian Noise (AWGN);
and  $\mathbf{h}_k(t)\in\mathbb{C}^{M\times 1}$ is the fading channel vector between the BS and user $k$. 
We refer to the power $P$ as the SNR.
Moreover, we make the widely adopted assumption that Channel State Information at the Receivers (CSIR) is perfect. 
On the other hand, CSIT is subject to uncertainty as explained in what follows.
\subsection{Channel State Information at the Transmitter}
\label{sec: CSIT}
For  each user $k$ and channel use $t$, the BS obtains a possibly imperfect estimate of the channel vector  
$\mathbf{h}_k(t)$, denoted by $\widehat{\mathbf{h}}_k(t)$.
CSIT imperfection is modelled by 
\begin{equation}
\label{eq: imperfectCSIT}
\mathbf{h}_k(t)=\widehat{\mathbf{h}}_k(t)+\widetilde{\mathbf{h}}_k(t),
\end{equation}
where $\widetilde{\mathbf{h}}_k(t)$ denotes the corresponding channel estimation error at the BS.
For compactness, we define $\mathbf{H}(t)\triangleq[\mathbf{h}_1(t) \cdots \mathbf{h}_K(t)]$, $\widehat{\mathbf{H}}(t)\triangleq[\widehat{\mathbf{h}}_1(t) \cdots \widehat{\mathbf{h}}_{K}(t)]$ and 
$\widetilde{\mathbf{H}}(t)\triangleq[\widetilde{\mathbf{h}}_1(t) \cdots \widetilde{\mathbf{h}}_K(t)]$, from which we have 
$\mathbf{H}(t)=\widehat{\mathbf{H}}(t)+\widetilde{\mathbf{H}}(t)$. 
For simplicity, we assume that the pairs $\big( \mathbf{H}(t),\widehat{\mathbf{H}}(t) \big)$ 
are i.i.d. over the time index $t$.
This implies that the  joint process $\big\{ \mathbf{H}(t),\widehat{\mathbf{H}}(t)\big\}$
is stationary ergodic, which allows us to drop $t$ and replace time 
averages by ensemble averages in the regime $T \rightarrow \infty$.
The joint distribution $f_{\mathbf{H},\widehat{\mathbf{H}}}(\mathbf{H},\widehat{\mathbf{H}})$ 
is continuous and known to the BS.

For each user $k$, we define the average channel (power) gain as $G_{k} \triangleq \mathbb{E}\big\{\| \mathbf{h}_k \|^{2} \big\} $.
Similarly, we define $\widehat{G}_{k} \triangleq \mathbb{E}\big\{\| \widehat{\mathbf{h}}_k \|^{2} \big\} $ and 
$\widetilde{G}_{k} \triangleq \mathbb{E}\big\{\| \widetilde{\mathbf{h}}_k \|^{2} \big\} $. 
For many CSIT acquisition mechanisms, $\widehat{\mathbf{h}}_k$
and $\widetilde{\mathbf{h}}_k$ are uncorrelated.
By further assuming that $\widehat{\mathbf{h}}_k$ and $\widetilde{\mathbf{h}}_k$ have zero means, 
we have $G_{k} =  \widehat{G}_{k}  + \widetilde{G}_{k} $, from which we can write 
$\widehat{G}_{k} = (1 - \sigma_{e,k}^2) G_{k}$ and $\widetilde{G}_{k} = \sigma_{e,k}^2 G_{k}$ for some $\sigma_{e,k}^2\in[0,1]$.
Note that $\sigma_{e,k}^2 $ is the
normalized estimation error variance for user $k$'s CSIT, e.g. $\sigma_{e,k}^2=1$ represents no instantaneous CSIT,
while  $\sigma_{e,k}^2=0$ represents perfect instantaneous CSIT. 
\subsection{Heterogeneous CSIT and Receiver Capability}
\label{subsec:user_sets}
We consider settings where users scheduled by the BS fall into one of two categories: 
$\alpha$-users that provide instantaneous CSIT feedback and can 
benefit from spatial multiplexing gains, and $0$-users that do not provide instantaneous CSIT feedback.
These are further described as follows.
\begin{itemize}
\item \textit{$\alpha$-users}:  
The BS schedules $M$ such users, indexed by $\mathcal{K}_{\alpha} \triangleq [1 : M]$.
We consider an equal normalized CSIT error variance amongst these users, i.e. $\sigma_{e,k}^2 = \sigma_{e}^2$
for all $k  \in \mathcal{K}_{\alpha}$.
To facilitate the DoF analysis in Section \ref{sec: DoF}, 
we assume that $\sigma_{e}^2 $ scales with SNR  as $ \sigma_{e}^2 = P^{-\alpha}$ for some CSIT quality parameter $\alpha \in [0, \infty)$, where the two  extremes $\alpha = 0$ and $\alpha = \infty$ correspond to 
no CSIT and perfect CSIT, respectively.
As far as the DoF analysis is concerned, however, we may truncate
the CSIT quality parameters as $\alpha\in[0,1]$, where 
$\alpha = 1$ amounts to perfect CSIT in the DoF sense \cite{AG2015}.
The regime $\alpha \in (0,1)$ corresponds to partial CSIT, resulting from imperfections in acquisition (e.g. limited feedback).
\item \textit{$0$-users}: These are indexed by $\mathcal{K}_{0}=[ M+1 : K]$, and do not provide instantaneous
CSIT feedback. 
Nevertheless, the BS has access to statistical CSI (i.e. long-term CSI) for these users, based on the knowledge of $f_{\mathbf{H},\widehat{\mathbf{H}}}(\mathbf{H},\widehat{\mathbf{H}})$.
The above implies that  $\sigma_{e,k}^2 =1 $, $\widehat{\mathbf{h}}_k= \mathbf{0}$ and $\mathbf{h}_k=\widetilde{\mathbf{h}}_k$, for all $k\in\mathcal{K}_{0}$.
Note that the terminology $0$-users refers to the fact that these users have a CSIT quality parameter of zero, i.e. $\sigma_{e,k}^2 = P^{-0} $ for all 
$k\in\mathcal{K}_{0}$.
\end{itemize}
From the above, it is evident that  $\mathcal{K}=\mathcal{K}_{\alpha}\cup\mathcal{K}_{0}$.
The above categorization is motivated by cellular IoT networks, where $\alpha$-users represent high-end receivers (e.g. smart phones) with sophisticated functions; while  $0$-users represent low-end receivers (e.g. IoT devices) with simple functions.
This heterogeneous receiver capability will also influence our design as we will see further on, where we focus on schemes in which $\alpha$-users can implement successive decoding, while  $0$-users are limited to simple decoding only.
\vspace{-0.5cm}
\subsection{Messages, Codewords, Rates and DoF}
\label{sec: def DoF}
The BS has the messages $W_1,\ldots,W_K$, each intended to its respective user. 
In a communication session of $T$ channel uses, messages are mapped to a codeword $\{ \mathbf{x}(t) \} \triangleq \{ \mathbf{x}(1) \cdots
\mathbf{x}(T) \}$, which satisfies the aforementioned average power constraint. 
This mapping may depend on the estimates $ \{  \widehat{\mathbf{H}}(t) \}$ (instantaneous CSIT) and the distribution 
$f_{\mathbf{H},\widehat{\mathbf{H}}}(\mathbf{H},\widehat{\mathbf{H}})$  (long-term CSIT), but not on the exact channel realizations
$\{  \mathbf{H}(t) \}$, which are unknown to the BS.
Achievable rates are defined in the ergodic sense in a standard Shannon theoretic fashion, where 
$(\overline{R}_1(P),\cdots,\overline{R}_K(P))$ denotes an achievable Ergodic Rate (ER) tuple for a given SNR,
and $\mathcal{C}(P)$ denotes the capacity region.
A DoF tuple $(d_1, \cdots, d_K)$ is said to be achievable if there exists $(\overline{R}_1(P),\cdots,\overline{R}_K(P))  \in \mathcal{C}(P)$, for all $P > 0$, such that $d_k\triangleq\lim_{P \to \infty} \frac{\overline{R}_k(P)}{\log(P)}$ for all $k \in \mathcal{K}$.
The DoF region  $\mathcal{D}$ is defined as the closure of all achievable DoF tuples.  
\section{Rate-Splitting Multiple Access}
\label{sec: RSMA}
The capacity region of the MISO BC with partial CSIT is unknown in general.
For underloaded settings, i.e. $K \leq M$, however, the DoF region has been fully characterized and shown to be achievable using RS-based schemes \cite{enrico2017bruno,hamdi2019spawc}.
The optimality of RS in the asymptotically high-SNR regime in underloaded settings, shown through DoF analysis, provides firm theoretical grounds for further design and optimization in the finite SNR regime, see, e.g., \cite{RS2016hamdi,mao2019beyondDPC}.

Moving back to the overloaded setting of interest, i.e. $K > M$,  the DoF region in this case is not known, and the results in 
\cite{enrico2017bruno,hamdi2019spawc} do not directly extend to this setting.
Nevertheless, motivated by the DoF optimality of RS in underloaded settings, in this section we propose two schemes that generalize RSMA to the overloaded MISO BC of interest, namely \textit{Time Partitioning} (TP)--RSMA and  \textit{Power Partitioning} (PP)--RSMA.

\vspace{-0.4cm}
\subsection{Time Partitioning (TP)--RSMA}
\vspace{-0.1cm}
\label{sec: TPRS}
\begin{figure}[t!]
	\centering
	\includegraphics[width=4.5in]{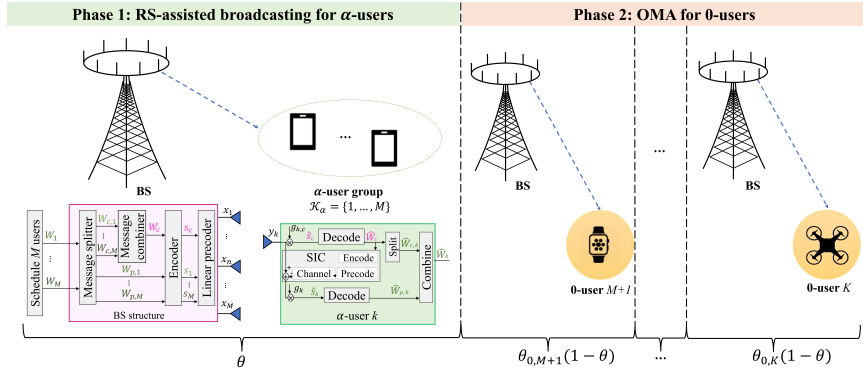}%
		\vspace{-1mm}
	\caption{TP-RSMA-assisted $K$-user overloaded MISO BC with heterogeneous CSIT.}
	\label{fig: systemModelTPRS}
		\vspace{-8mm}
\end{figure}
A natural strategy is to treat the two sets of users $\mathcal{K}_{\alpha}$ and $\mathcal{K}_{0}$ separately.
The set of $\alpha$-users in isolation constitutes an underloaded MISO BC with partial CSIT, for which 
RS is DoF-optimal \cite{enrico2017bruno,hamdi2019spawc}.
On the other hand, for a setting comprising $0$-users on their own, 
simple OMA  is DoF-optimal, as spatial multiplexing gains collapse due to the absence of CSIT \cite{AG2015}.    
Therefore, the communication session of $T$ channel uses can be partitioned in time into two phases, given some 
TP factor $\theta\in[0,1]$: one occupying $\theta T$ channel uses where $\mathcal{K}_{\alpha}$  are served using RS,
and another occupying the remaining $(1- \theta) T$ channel uses in which $\mathcal{K}_{0}$ are served using OMA.
This is illustrated in Fig. \ref{fig: systemModelTPRS}.
\subsubsection{Phase 1}
In this phase, users in $\mathcal{K}_{\alpha}$ are served using the linearly-precoded RS scheme  in  \cite{RS2016hamdi}, which is described as follows.
For each user $k$ in $\mathcal{K}_{\alpha}$, the corresponding message $W_{k}$ is split into a 
common part $W_{c,k}$ and a private part $W_{p,k}$.
The common parts are combined into one common message $W_c$.
The $M+1$ resulting messages $W_c,W_{p,1},\cdots,W_{p,M}$ are respectively encoded into the independent Gaussian data streams 
$\{s_c(t)\},\{s_1(t)\},\cdots,\{s_M(t)\}$, each of length $\theta T$.
Focusing on an arbitrary channel use and dropping the time index $t$, the $M+1$
data streams, given by the vector $\mathbf{s}_{\alpha}\triangleq[s_c \; s_1 \; \cdots \; s_M]^{\Trn}$, are linearly precoded using
the precoding matrix 
$\mathbf{P}_{\alpha}\triangleq[\mathbf{p}_c \; \mathbf{p}_1 \; \cdots \; \mathbf{p}_M]\in\mathbb{C}^{M\times (M+1)}$. The transmit signal becomes
\begin{equation}
\label{eq: transmit signal TPRS}
\mathbf{x}=\mathbf{P}_{\alpha}\mathbf{s}_{\alpha}={{\mathbf{p}_c{s}_c}}+{{\sum_{k\in\mathcal{K}_{\alpha}}\mathbf{p}_{k}{s}_{k}}}.
\end{equation}
Note that in the above, we have $\mathbb{E}\{\mathbf{s}_{\alpha}\mathbf{s}_{\alpha}^{\Hrm}\}=\mathbf{I}$ and $\mathrm{tr}(\mathbf{P}_{\alpha}\mathbf{P}_{\alpha}^{\Hrm}) \leq P$.

On the receiver side, each user $k$ in $\mathcal{K}_{\alpha}$ decodes both $s_c$ and $s_k$ using successive decoding.
In particular, $s_c$ is decoded first by treating interference from all other streams as Gaussian noise, from which user $k$
recovers the desired common part $W_{c,k}$.
Then  $s_c$  is cancelled from the received signal and user $k$ proceeds to decode $s_k$, from which  the desired private  part $W_{p,k}$ is recovered. From $W_{c,k}$ and $W_{p,k}$, the desired message $W_k$ is retrieved.
Consider a given precoding scheme $\mathbf{P}_{\alpha}$ and let us assume that $\theta = 1$, which will be relaxed further on. 
The instantaneous Signal-to-Interference-plus-Noise Ratio (SINR) of $W_c$ and $W_{p,k}$ are given as
\begin{equation}
\label{eq: SINR cp}
\gamma_{c,k}=\frac{|{\mathbf{h}}_{k}^{\Hrm}\mathbf{p}_{c}|^2}{\sum\limits_{j\in\mathcal{K}_{\alpha}}|\mathbf{h}_{k}^{\Hrm}\mathbf{p}_{j}|^2+1},
\quad
\gamma_{p,k}=\frac{|{\mathbf{h}}_{k}^{\Hrm}\mathbf{p}_{k}|^2}{\sum\limits_{j\in\mathcal{K}_{\alpha}\setminus k}|\mathbf{h}_{k}^{\Hrm}\mathbf{p}_{j}|^2+1}.
\end{equation}

\begin{equation}
\label{eq: instantanesous}
\overline{R}_{c,k} = \mathbb{E}_{\{ \mathbf{H},\widehat{\mathbf{H}}\}} \left\{ \log\left(1+\gamma_{c,k}\right) \right\},
\quad
\overline{R}_{p,k} = \mathbb{E}_{\{ \mathbf{H},\widehat{\mathbf{H}}\}} \left\{ \log\left(1+\gamma_{p,k}\right) \right\},
\end{equation}
respectively. The above expectations are taken with respect to the joint process $\{ \mathbf{H},\widehat{\mathbf{H}}\}$.
To further guarantee that $W_c$ is successfully decoded by all users, its ER
 $\overline{R}_{c}$ must not exceed $ \min\{ \overline{R}_{c,k} \mid k\in \mathcal{K}_{\alpha}\}$.
The rate $\overline{R}_{c} $ is shared across the $M$ users in $\mathcal{K}_{\alpha}$, where each 
user $k$  is allocated a portion $\overline{C}_{k}$ corresponding to the rate of $W_{c,k}$,
such that $\sum_{k \in \mathcal{K}_{\alpha}} \overline{C}_{k} = \overline{R}_{c} $.

It follows from the above that for any $\theta \in [0,1]$, the ER achieved by each user $k$ in $\mathcal{K}_{\alpha}$ 
is given by the sum of two contributions, i.e. common and private, as follows
\begin{equation}
\vspace{-0.2cm}
\overline{R}_{k}^{\textrm{TP}}  = \theta \big( \overline{C}_{k} + \overline{R}_{p,k} \big) , \ \forall k \in \mathcal{K}_{\alpha}.
\end{equation}
\subsubsection{Phase 2}
Users in $\mathcal{K}_{0}$ are served in an OMA fashion,
where each user $k$ in $\mathcal{K}_{0}$ is allocated a fraction $\theta_{0,k}$ 
of this phase's duration, 
such that $\sum_{k \in \mathcal{K}_{0}} \theta_{0,k} = 1$. 
Without loss of generality, we assume uniform time allocation among $0$-users, i.e.,  $\theta_{0,k}=\frac{1}{K-M}$ in this work. 
The signal dedicated to each user $k$ in $\mathcal{K}_{0}$, transmitted during the designated  $\theta_{0,k} (1 - \theta) T$ channel uses, is given by 
\vspace{-0.2cm}
\begin{equation}
\label{eq: transmit signal TPRS_2}
\mathbf{x}=\mathbf{p}_{0}{s}_k,
\vspace{-0.3cm}
\end{equation}
where $s_k$ is the corresponding data stream of user $k$ in $\mathcal{K}_{0}$ satisfying $\mathbb{E}\{|s_k|^{2}\} = 1$, while 
$\mathbf{p}_{0}$ is a precoding vector satisfying
$\| \mathbf{p}_{0} \|^{2} \leq P$.
As seen from \eqref{eq: transmit signal TPRS_2}, we assume that the same precoding vector is used across all $0$-users.
While this assumption is mainly motivated by simplicity, no loss of optimality is incurred whenever $0$-users have similar channel statistics (i.e. statistically equivalent).
The ER achieved by each user $k$ in $\mathcal{K}_{0}$  is hence given by
\begin{equation}
\label{eq: noCSITRate TPRS}
\overline{R}_{k}^{\textrm{TP}}= \theta_{0,k}   (1-\theta) \mathbb{E}_{\mathbf{H}}  \left\{ \log\left(1+\gamma_{0,k}^{\textrm{TP}}\right)
 \right\}, \; \forall k \in \mathcal{K}_0.
\end{equation}
where $\gamma_{0,k}^{\textrm{TP}}={|{\mathbf{h}}_{k}^{\Hrm}\mathbf{p}_{0}|^2}$ is the instantaneous SNR of decoding $s_k$ at the corresponding user.
\subsection{Power Partitioning (PP)--RSMA}
\label{sec: PPRS}
\begin{figure}[t!]
	\centering
	\includegraphics[width=6in]{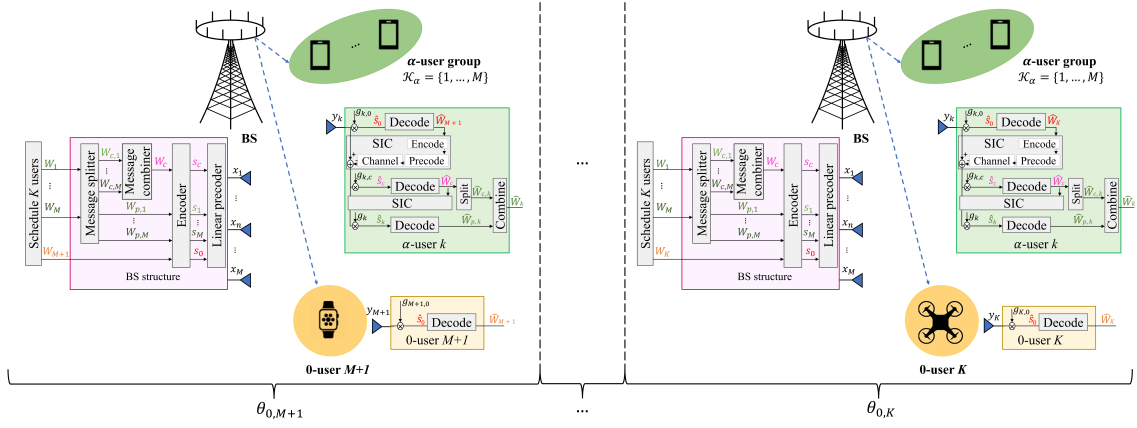}%
		\vspace{-1mm}
	\caption{PP--RSMA-assisted $K$-user overloaded MISO BC with heterogeneous CSIT.}
	\label{fig: systemModelPPRS}
		\vspace{-8mm}
\end{figure}

Contrary to TP--RSMA,  in PP--RSMA we propose to treat the two sets of users $\mathcal{K}_{\alpha}$ and $\mathcal{K}_{0}$ jointly.
In particular, the transmit signal is composed by superimposing a RS signal intended to $\alpha$-users, i.e. similar to the one in
 \eqref{eq: transmit signal TPRS}, and an OMA signal intended to $0$-user, i.e. similar to the one in \eqref{eq: transmit signal TPRS_2},
 yielding
\begin{equation}
\label{eq: transmit signal DPCRS}
\mathbf{x}=\mathbf{P}\mathbf{s}={{\mathbf{p}_0{s}_0}}+{{\mathbf{p}_c{s}_c}}+{{\sum_{k\in\mathcal{K}_{\alpha}}\mathbf{p}_{k}{s}_{k}}}
\vspace{-1mm}
\end{equation}
where $\mathbf{P}\triangleq[\mathbf{p}_0 \; \mathbf{P}_{\alpha}] \in\mathbb{C}^{M\times (M+2)}$ and  
$\mathbf{s}\triangleq[s_0 \; \mathbf{s}_{\alpha}^{\Trn}]^{\Trn} \in\mathbb{C}^{(M+2)\times 1}$. 
The stream $s_0$  in \eqref{eq: transmit signal DPCRS} is time-shared amongst $0$-users,
such that each user $k$ in $\mathcal{K}_{0}$ is allocated a fraction $\theta_{0,k}$ of the total duration.
Moreover, we have $\mathbb{E}\{\mathbf{s}\mathbf{s}^{\Hrm}\}=\mathbf{I}$ and 
$\mathrm{tr}(\mathbf{P}\mathbf{P}^{\Hrm}) = \| \mathbf{p}_{0} \|^{2} + 
\mathrm{tr}(\mathbf{P}_{\alpha}\mathbf{P}_{\alpha}^{\Hrm}) \leq P$,
where the latter indicates that the total transmit power is partitioned between $0$-users and $\alpha$-users, 
hence the PP terminology.
This scheme is illustrated in Fig. \ref{fig: systemModelPPRS}.

At the receiver side, each $0$-user $k$ in $\mathcal{K}_{0}$ decodes the designated portion of $s_{0}$, occupying $\theta_{0,k} T$ channel uses,  while treating interference from all other streams as noise. 
On the other hand, each $\alpha$-user $k$ in $\mathcal{K}_{\alpha}$ first decodes and cancels the stream  $s_{0}$, and then proceeds to decode $s_{c}$ and $s_{k}$ as in the regular RS fashion described as part of the TP--RSMA scheme.
It follows that for any given precoding scheme $\mathbf{P}$,
an $\alpha$-user $k$ achieves an ER of
\begin{equation}
\vspace{-0.2cm}
\overline{R}_{k}^{\textrm{PP}}  =  \overline{C}_{k} + \overline{R}_{p,k}  , \ \forall k \in \mathcal{K}_{\alpha}.
\end{equation}
where $\sum_{k \in \mathcal{K}_{\alpha}} \overline{C}_{k} = \overline{R}_{c} = \min\{ \overline{R}_{c,k} \mid k\in \mathcal{K}_{\alpha}\}$, and  $\overline{R}_{c,k}, \overline{R}_{k}$ are defined as in \eqref{eq: instantanesous}.
Now we recall that the signal intended to each $0$-user $k$ in $\mathcal{K}_{0}$, occupying a portion $\theta_{0,k}$ of $s_{0}$, is decoded by all users in $ \mathcal{K}_{\alpha} \cup \{k\}$.
The instantaneous SINR of decoding $s_0$ at each user is given as
\begin{equation}
\label{eq: noCSITSINR PP}
\gamma_{0,k}^{\textrm{PP}} = \frac{|{\mathbf{h}}_{k}^{\Hrm}\mathbf{p}_{0}|^2}{|\mathbf{h}_{k}^{\Hrm}\mathbf{p}_{c}|^2+\sum_{j\in\mathcal{K}_{\alpha}}|\mathbf{h}_{k}^{\Hrm}\mathbf{p}_{j}|^2+1}, \ \forall k \in 
\mathcal{K}_{0}.
\end{equation}
Each such user achieves an ER of 
\begin{equation}
\label{eq: noCSITRate}
\overline{R}_{k}^{\textrm{PP}} = \theta_{0,k} \min_{k \in \mathcal{K}_{\alpha} \cup \{k\} } \left\{  \mathbb{E}_{\{ \mathbf{H},\widehat{\mathbf{H}}\}}  \left\{ \log\left(1+\gamma_{0,k}^{\textrm{PP}}\right) \right\} 
\right\}, \ \forall k \in 
\mathcal{K}_{0},
\end{equation}
which guarantees successful decoding of corresponding messages. Before we proceed, 
it is worth noting that the PP--RSMA scheme leverages the layered superposition nature of the RS scheme, and incorporates the $0$-users transmission as an additional layer within the RS framework.

\begin{remark}
Notice that besides the proposed PP--RSMA scheme, the PP--SDMA scheme, which leverages MU--LP based SDMA  and incorporates the 0-users transmission as an additional layer within the SDMA framework, has not been investigated yet. It is another scheme we proposed in this work. Since PP--SDMA is a subset of PP--RSMA by turning off the common stream (i.e.,  forcing $\|\mathbf{p}_c\|^2=0$), the  system model of PP--SDMA is not described exclusively. More details of PP--SDMA are presented in Section \ref{sec: numerical results}.
\end{remark}


\vspace{-0.4cm}
\section{DoF Analysis}
\label{sec: DoF}
Having proposed two strategies for the overloaded MISO BC, in this section we analytically study and compare their performances. Achievable ERs, however, are highly coupled with design variables (i.e. precoding vectors and partition factors), which require non-trivial and intricate optimization, as we will see further on. Therefore, it is generally not possible to derive closed-form expressions of the ERs suitable for analysis and gaining insights. As an alternative, we resort to DoF analysis. In this context, the DoF can be thought of as a first-order approximation of the ER in the asymptotically high SNR regime (i.e. the interference limited regime). 
We characterize the DoF performances of TP-RSMA and PP-RSMA, from which we show that the latter achieves strictly superior performances. We then prove that PP-RSMA is DoF optimal for the considered setting, i.e. it achieves the entire optimal DoF region of the overloaded MISO BC. This settles the DoF region problem for this channel.


\subsection{Achievable DoF}
\label{sec: Achievable DoF}
One advantage of DoF analysis is that it is sufficient to consider low-complexity precoders, see, e.g., \cite{RS2016hamdi,hamdi2017bruno,hamdi2016robust}. With this in mind, we use random precoding vectors for  $\mathbf{p}_0$ and $\mathbf{p}_c$, i.e., $\mathbf{p}_0=\mathbf{e}_1\|\mathbf{p}_0\|^2, \mathbf{p}_c=\mathbf{e}_1\|\mathbf{p}_c\|^2$  (where $\mathbf{e}_1$ is a null vector with one entry to be equal to 1 and all other entries to be 0) and Zero-Forcing (ZF)  precoding vectors for $\{\mathbf{p}_k| k\in\mathcal{K}_{\alpha}\}$, designed using the channel estimate such that $\mathbf{p}_k\perp\{\widehat{\mathbf{h}}_i\}_{i\in\mathcal{K}_{\alpha}\setminus k}$. Hence, the residual interference power from the private streams of other $\alpha$-users $\sum_{i\in\mathcal{K}_{\alpha}\setminus k}|\mathbf{h}_{k}^{\Hrm}\mathbf{p}_{i}|^2$ received at $\alpha$-user $k$  reduces to $\sum_{i\in\mathcal{K}_{\alpha}\setminus k}|\widetilde{\mathbf{h}}_{k}^{\Hrm}\mathbf{p}_{i}|^2$.
The achievable DoF of TP--RSMA and PP--RSMA strategies are specified in the following.

\subsubsection{TP--RSMA}
In  phase 1 of TP--RSMA, $\alpha$-users are active and the power allocation is made to scale as\footnote{We use the standard Landau notation $\mathcal{O}(\cdot)$ (also known as ``big-O notation") to describe power scaling. Specifically, for real-valued functions $f(P)$ and $g(P)$, the statement  $f(P)=\mathcal{O}(g(P))$ means that $\lim_{P\to\infty}\frac{|f(P)|}{|g(P)|}<\infty$. }
\begin{equation}
\label{eq: PA TP}
\begin{aligned}
	\left \|\mathbf{p}_c  \right \|^2 &=\mathcal{O}(P)\\
	\left \|\mathbf{p}_k  \right \|^2 &=\mathcal{O}(P^{\alpha}), \,\,\forall k\in\mathcal{K}_{\alpha}.
\end{aligned}
\end{equation}
Recall that $\alpha$ corresponds to the CSIT quality parameter, and is in $[0,1]$. One example of a power allocation scheme that satisfies (\ref{eq: PA TP}) and the transmit power constraint is $\left \|\mathbf{p}_c  \right \|^2=P-P^{\alpha}$ and $\left \|\mathbf{p}_k  \right \|^2=\frac{P^{\alpha}}{M},\forall k\in\mathcal{K}_{\alpha}$. Each user first decodes the  common stream $s_c$ by treating all other streams as noise. Once the common stream  is decoded and removed from the received signal, each $\alpha$-user decodes the intended private stream. From the SINR expressions specified in (\ref{eq: SINR cp}) and the power allocation  in (\ref{eq: PA TP}), it follows that   at each $\alpha$-user $k$, the received SINRs of  the common stream and private stream scale as $\gamma_{c,k}=\mathcal{O}(P^{1-\alpha})$ and  $\gamma_{p,k}=\mathcal{O}(P^{\alpha})$, respectively\footnote{ Without loss of generality, we assume the channel (power) gain of each user  $\| \mathbf{h}_k \|^{2}$ is bounded above and below by positive constants, i.e., $\| \mathbf{h}_k \|^{2}\in[\Delta_1,\Delta_2]$. As $|{\mathbf{h}}_{k}^{\Hrm}\mathbf{p}_{c}|^2=|{\mathbf{h}}_{k}^{\Hrm}\mathbf{e}_{1}|^2\|\mathbf{p}_{c}\|^2\geq \Delta_1\|\mathbf{p}_{c}\|^2$, $|{\mathbf{h}}_{k}^{\Hrm}\mathbf{p}_{k}|^2\leq \|{\mathbf{h}}_{k}\|^2\|\mathbf{p}_{k}\|^2= \Delta_2\|\mathbf{p}_{k}\|^2$,
and $\sum_{i\in\mathcal{K}_{\alpha}\setminus k}|\mathbf{h}_{k}^{\Hrm}\mathbf{p}_{i}|^2+1= \sum_{i\in\mathcal{K}_{\alpha}\setminus k}|\widetilde{\mathbf{h}}_{k}^{\Hrm}\mathbf{p}_{i}|^2+1\leq\sigma_{e}^2\Delta_2\sum_{i\in\mathcal{K}_{\alpha}\setminus k}\|\mathbf{p}_{i}\|^2+1$, following (\ref{eq: SINR cp}), we therefore obtain that   SINR $\gamma_{c,k}\geq \frac{\Delta_1\|\mathbf{p}_{c}\|^2}{\Delta_2\|\mathbf{p}_{k}\|^2+\sigma_{e}^2\Delta_2\sum_{i\in\mathcal{K}_{\alpha}\setminus k}\|\mathbf{p}_{i}\|^2+1}$   scales as $\mathcal{O}(P^{1-\alpha})$. Similarly, we could  obtain that  SINR $\gamma_{k}$ scales as $\mathcal{O}(P^{\alpha})$.}. Hence, after  normalizing by the TP factor $\theta$, the  DoF achieved by  common stream $s_c$ is $1-\alpha$ while the  DoF achieved by each private stream is $\alpha$. By evenly sharing the common stream among $\alpha$-users, the per user symmetric normalized DoF   is the sum of evenly allocated DoF of the common stream $\frac{1-\alpha}{M}$ and the DoF of its private stream $\alpha$, which is equivalent to $\frac{1+(M-1)\alpha}{M}$. 

In  phase 2 of TP--RSMA,  $0$-users are served in an orthogonal fashion with full power allocated to the  stream $s_k$ for a fraction $\theta_{0,k}$ 
of the duration of phase 2. The SNR $\gamma_{0,k}^{\textrm{TP}}$ of a $0$-user $k$   scales as $\mathcal{O}(P)$. The DoF therefore collapses to 1 after normalizing by the time duration $\theta_{0,k}(1-\theta)$.   As uniform time allocation among $0$-users is assumed in this work, i.e., $\theta_{0,k}=\frac{1}{K-M}$, the  DoF of each user is  $\frac{1}{K-M}$ after normalizing by the time partition $1-\theta$.

By incorporating the TP factor $\theta\in[0,1]$, the actual DoF $d_k$ achieved by  user $k$ using TP--RSMA is given by
\begin{equation}
\label{eq: DoF TPRS}
d_k^{\textrm{TP}}=\left\{\begin{matrix}
\theta\frac{1+(M-1)\alpha}{M},& k\in\mathcal{K}_{\alpha} \\ 
(1-\theta)\frac{1}{K-M}, & k\in\mathcal{K}_{0}
\end{matrix}\right.
\end{equation}

\subsubsection{PP--RSMA}
In order to partition the signal-space through the power domain, the power allocation of the PP--RSMA scheme is carried out such that 
\begin{equation}
\label{eq: PA PP}
\begin{aligned}
&\left \|\mathbf{p}_0  \right \|^2 =\mathcal{O}(P)\\
&\left \|\mathbf{p}_c  \right \|^2+\sum_{k\in\mathcal{K}_{\alpha}}\left \|\mathbf{p}_k  \right \|^2=\mathcal{O}(P^{\beta}),
\end{aligned}
\end{equation}
where $\beta\in [0,1]$ is the PP factor. The bottom $\beta$ power levels are reserved for the transmission to $\alpha$-users while the top $1-\beta$ power levels are occupied by the transmission to $0$-users. 

At the receiver side, each $0$-user directly decodes the common stream $s_0$ by treating interference from streams intended to $\alpha$-users as noise. The SINR of each $0$-user to decode $s_0$ given in (\ref{eq: noCSITSINR PP}) scales as $\mathcal{O}(P^{1-\beta})$. The  DoF achieved by $0$-users is $1-\beta$ after normalizing by the time duration $\theta_{0,k}$. As $\theta_{0,k}=\frac{1}{K-M}$, the  DoF of each $0$-user is  $\frac{1-\beta}{K-M}$.
For $\alpha$-users, the power of $\mathcal{O}(P^{\beta})$  is further split as 
\begin{equation}
\label{eq: PA PP 2}
\begin{aligned}
\left \|\mathbf{p}_c  \right \|^2 &=\mathcal{O}(P^{\beta})\\
\left \|\mathbf{p}_k  \right \|^2 &=\mathcal{O}(P^{\tau}), \,\,\forall k\in\mathcal{K}_{\alpha},
\end{aligned}
\end{equation} 
where $\tau\leq \beta$. One example power allocation scheme that satisfies (\ref{eq: PA PP}), (\ref{eq: PA PP 2}) and the transmit power constraint is $\left \|\mathbf{p}_0  \right \|^2=P-P^{\beta}$, $\left \|\mathbf{p}_c  \right \|^2=P^{\beta}-P^{\tau}$ and $\left \|\mathbf{p}_k  \right \|^2=\frac{P^{\tau}}{M}, \forall k\in\mathcal{K}_{\alpha}$.
The respective SINRs of each $\alpha$-user to decode the intended common and private streams scale as $\mathcal{O}(P^{\beta-\tau})$ and $\mathcal{O}(P^{\tau})$, which follows  the SINR specified in (\ref{eq: SINR cp}). Hence, the  DoF achieved by the common stream  is $\beta-\tau$ and that achieved by each private stream is $\tau$.

 It remains to highlight that as the channel estimation error $\sigma_{e,k}^2$ scales as $\mathcal{O}(P^{-\alpha})$, the interference experienced at each $\alpha$-user scales as  $\mathcal{O}(P^{\tau-\alpha})$ under the assumption of ZF precoding. If $\tau\leq\alpha$, the interference is drowned by noise. Knowing that $\tau\leq \beta$, we can set  $\tau=\min\{\alpha,\beta\}$. When $\beta\leq\alpha$, ZF is sufficient for  $\alpha$-users to reduce the interference to the power level of noise and RS is unnecessary (i.e. there is no benefit DoF wise to allocate any power to the common stream $s_c$). When $\beta>\alpha$, each $\alpha$-user needs to rely on RS to manage interference.
 
  Based on the above analysis, we obtain that each private stream achieves a DoF of $\min\{\alpha,\beta\}$ while the common stream achieves a DoF of $\beta-\min\{\alpha,\beta\}$.  By evenly allocating the DoF of the common stream among users, the symmetric DoF achieved by user $k$ using PP--RSMA is
\begin{equation}
\label{eq: DoF PPRS}
d_k^{\textrm{PP}}=\left\{\begin{matrix}
\frac{\beta+(M-1)\min\{\alpha,\beta\}}{M},& k\in\mathcal{K}_{\alpha} \\ 
\frac{1-\beta}{K-M}, & k\in\mathcal{K}_{0}
\end{matrix}\right.
\end{equation}

\begin{figure}
	\centering
	\begin{subfigure}[b]{0.35\textwidth}
		\vspace{-1mm}
		\centering
		\includegraphics[width=0.9\textwidth]{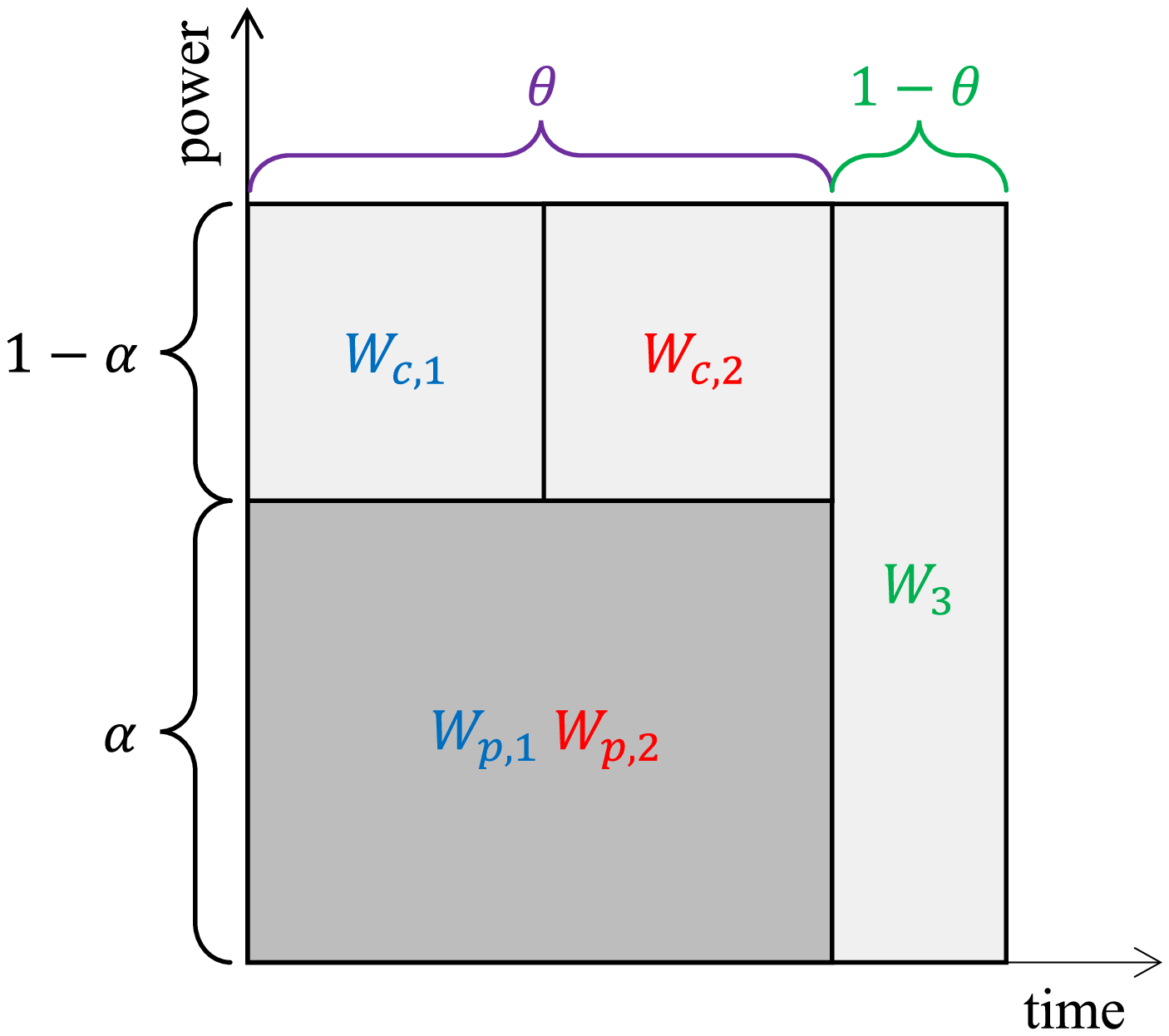}%
		\caption{TP--RSMA}
	\end{subfigure}%
	~
	\begin{subfigure}[b]{0.40\textwidth}
		\vspace{-1mm}
		\centering
		\includegraphics[width=0.9\textwidth]{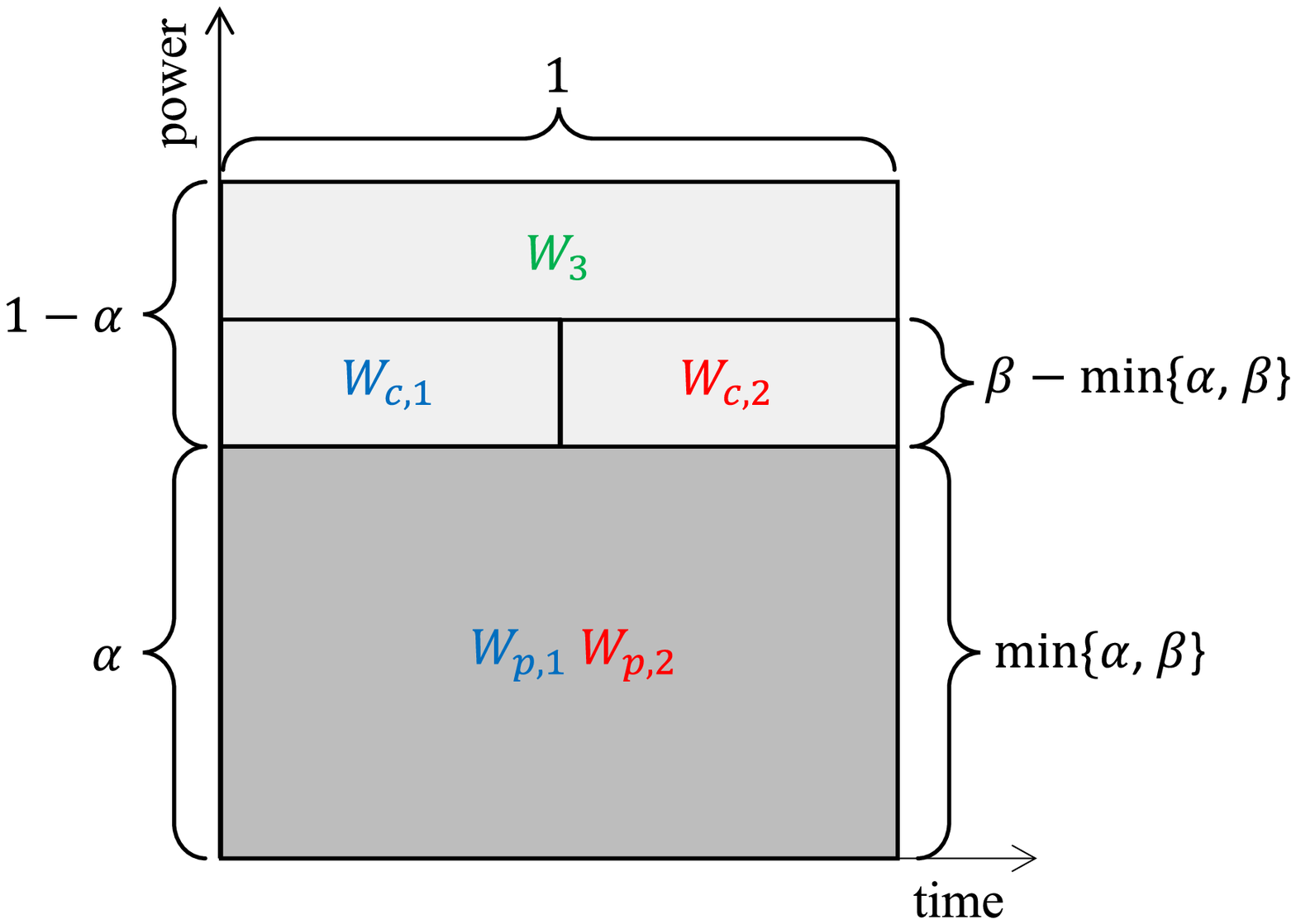}%
		\caption{PP--RSMA}
	\end{subfigure}%
	\vspace{-1mm}
	\caption{TP--RSMA and PP--RSMA for $M=2$ and $K=3$. Define the normalized spatial-multiplexing as the sum-DoF normalized by both TP--RSMA and PP--RSMA. The normalized spatial-multiplexing gain in rectangles with light and dark shadings is $1$ and $2$ respectively.}
	\label{fig: dof diagram}
	\vspace{-7mm}
\end{figure}

Fig. \ref{fig: dof diagram} illustrates an example of TP--RSMA and PP--RSMA strategies specified in this section. 

\subsubsection{Gain of PP--RSMA over TP--RSMA}
Next, we demonstrate the DoF gain of PP--RSMA over TP--RSMA. Symmetric DoFs achieved by $\alpha$-users using TP--RSMA and PP--RSMA are compared given that $0$-users maintain the equal DoF in both transmission strategies, i.e., $d_k^{\textrm{TP}}=d_k^{\textrm{PP}}, \forall k\in\mathcal{K}_0$. To achieve this, we have $\beta=\theta$ according to equation (\ref{eq: DoF TPRS}) and (\ref{eq: DoF PPRS}). Hence, the DoF  achieved by $\alpha$-user $k$ using PP--RSMA is 
\begin{equation}
\label{eq: DoF PPRS 2}
d_k^{\textrm{PP}}=\left\{\begin{matrix}
\frac{\theta+(M-1)\alpha}{M},& \alpha\leq \theta \\ 
\theta, & \alpha\geq \theta
\end{matrix}\right. , k\in\mathcal{K}_{\alpha}.
\end{equation}
We obtain that $d_k^{\textrm{PP}}\geq d_k^{\textrm{TP}}, \forall k\in\mathcal{K}$ by comparing (\ref{eq: DoF TPRS}) and (\ref{eq: DoF PPRS 2}).  Specifically, it follows that for any $\alpha$ in $[0,1]$, we have
\begin{equation}
\frac{\theta+(M-1)\min\{\theta,\alpha\}}{M}\geq	\theta\frac{1+(M-1)\alpha}{M}, \forall \theta \in[0,1]
\end{equation} 
when comparing the DoF for $\alpha$-users in  (\ref{eq: DoF TPRS}) and (\ref{eq: DoF PPRS 2}). Moreover, when $0<\theta, \alpha<1$, this inequality is strict. In other words, under the condition of partial CSIT for $\alpha$-users and non-zero non-unity power partitioning, PP--RSMA achieves a strict DoF improvement for $\alpha$-users over TP--RSMA.

\begin{figure}[t!]
	\centering
	\includegraphics[trim=0mm 0mm 0mm 0mm,clip=true,width=0.50\textwidth]{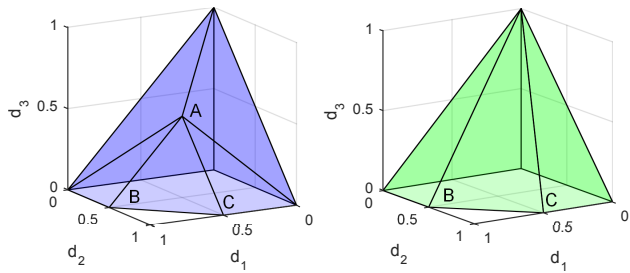}
	\caption{DoF region achieved by PP--RSMA (left) and TP--RSMA (right) for $M=2$ and $K=3$, and CSIT quality $\alpha=0.5$ for the first two users. The points are $A=(\alpha,\alpha,1-\alpha)$, $B=(1,\alpha,0)$ and $C=(\alpha,1,0)$. It can be seen that $A$ cannot be achieved through time partitioning.}
	\label{fig: dof region}
	\vspace{-0.7cm}
\end{figure}
With this example, we have shown the gain of PP--RSMA over TP--RSMA for a specific symmetric setup. While this provides an indication that PP--RSMA outperforms TP--RSMA, the question is whether this gain holds for general power and time allocation. To understand it, we need to characterize the achievable DoF region of both schemes. Surprisingly, we show that not only PP--RSMA outperforms TP--RSMA but in fact can achieve the optimal DoF region as we will show in Section \ref{sec: optimal DoF}. Before getting into the details,  we illustrate an example of the achievable DoF regions by the two schemes in Fig. \ref{fig: dof region} where the BS is equipped with  two transmit antennas and serves two $\alpha$-users (user 1 and user 2) and one $0$-user (user 3). In both subfigures, the DoF achieved in each rectangle (which is also known as a time-power resource block) is equal to the area of rectangle times the normalized spatial-multiplexing gain (2 for ZF and 1 for degraded). If user 3 is switched off, both TP--RSMA and PP--RSMA reduce to conventional RS and the sum DoF achieved by user 1 and user 2 is $1+\alpha$. Once user 3 is enabled, the sum DoF of TP--RSMA becomes $\theta(1+\alpha)+(1-\theta)=1+\theta\alpha$. In comparison, the sum DoF of PP--RSMA maintains $1+\alpha$  if $\beta=\theta\geq \alpha$ while the DoF becomes $\theta+\alpha+(1-\theta)=1+\theta$ if $\beta=\theta<\alpha$.   The sum DoF achieved by PP--RSMA is still higher than that  achieved by TP--RSMA.

\vspace{-3mm}
\subsection{Optimum DoF Region of Overloaded MISO BC}
\label{sec: optimal DoF}

The achievable DoF tuples in (\ref{eq: DoF TPRS}) and (\ref{eq: DoF PPRS 2}) are obtained under the assumption that random and ZF precoding is adopted at the BS,  $\alpha$-users achieve a symmetric DoF, i.e., $d_{k}=d_{\alpha},\forall k\in\mathcal{K}_{\alpha}$ and $0$-users achieve a symmetric DoF, i.e., $d_{k}=d_{0},\forall k\in\mathcal{K}_{0}$ for both TP--RSMA and PP--RSMA. Considering the symmetric DoF helps gain insights into the potential gains of PP-RSMA over TP-RSMA.  However, when considering more general scenarios, achievable DoF tuples assume a wide variety of tradeoffs characterized by achievable and optimum DoF regions. Interestingly, the optimum DoF region for the considered setup is achieved through variants of PP--RSMA, which is characterized in Theorem \ref{theorem 1}.
\vspace{-0.2cm}
\begin{theo} 
	\label{theorem 1}
	For the overloaded MISO BC described in Section \ref{sec: system_model}, the optimum DoF region $\mathcal{D}$ is given by
	\begin{equation} \label{dis_1}
	d_k \geq 0, \quad \forall k \in \mathcal{K}
	\end{equation}
	\begin{equation} \label{dis_2}
	\sum_{k \in \mathcal{S}}{d_{k}} +  \sum_{k \in \mathcal{K}_{\mathrm{0}}  }{d_k}  \leq 1 + (|\mathcal{S}|-1) \alpha, \quad \forall \mathcal{S} \subseteq  \mathcal{K}_{\mathrm{\alpha}}, |\mathcal{S}| \geq 1.
	\end{equation}
\end{theo}

The proof is relegated to Appendix I. We prove the optimality of $\mathcal{D}$ by showing that it is both achievable and an outer bound of the optimal region. The achievability of the DoF region is proved based on generalizing the PP--RSMA scheme of Section \ref{sec: PPRS} and the outer bound of the optimal region is proved based on the sum DoF upper bound in \cite{AG2015}.

To better visualize the optimum DoF region, an example is given in Fig. \ref{fig: dof region} (left) for a channel with $M=2$ and $K=3$, where the CSIT quality of the first two users is $\alpha=0.5$.
Moreover, for the sake of comparison, the DoF region achieved through TP--RSMA is shown in Fig. \ref{fig: dof region} (right).
The TP--RSMA region is obtained by time-sharing the DoF of 1 achieved by user 3 with the DoF region of the two remaining users achieved through RS (see \cite{RSintro16bruno}).
For the PP--RSMA region, the facet given by $A-B-C$ is in fact sum-DoF optimum. Hence, user 3 can be served with non-zero DoF without influencing the Sum-DoF (e.g. point $A$). On the other hand, serving user 3 with non-zero DoF through TP--RSMA is not possible without decreasing the sum-DoF as it requires moving away from the segment $B-C$.
\subsection{The Role of DoF Analysis}
\begin{figure}[t!]
	\centering
	\includegraphics[width=2.6in]{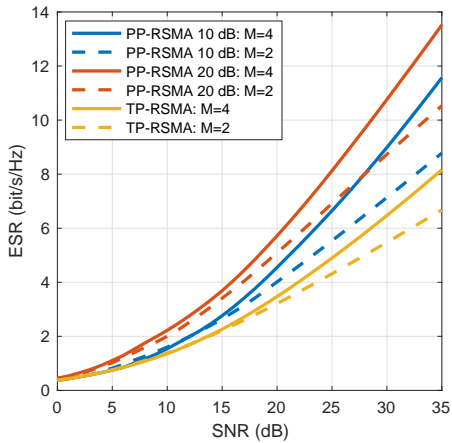}		
	\caption{Ergodic sum rate of $\alpha$-users based on random and ZF precoding. $K_0=60$, $\alpha=0.5$ and $\theta=0.5$.  Same rates for the $0$-users are maintained for the cases when the long-term SNR of the $0$-user is $10$ dB and $20$ dB lower than  that of $\alpha$-users.}
	\vspace{-6mm}
	\label{fig:Plot_1 a}
\end{figure}
Even though the low-complexity linear precoders discussed in \ref{sec: Achievable DoF} achieve the optimum DoF, they are suboptimal in the rate sense. Therefore, the question of how the DoF analysis in Section \ref{sec: DoF} affects the precoder design and the optimization process arises.

To highlight the potential role of DoF analysis in guiding the precoder design, we  illustrate in Fig. \ref{fig:Plot_1 a} the ESR  achieved by the DoF-motivated low-complex linear precoding scheme where  precoders $\mathbf{p}_0, \mathbf{p}_c$ of the common streams  are chosen as  random precoding vectors while the precoders of the private streams for $\alpha$-users are designed by ZF over the channel estimates. We consider two different settings with a massive connection to $0$-users being guaranteed. In the first setting, 
 the BS has two  transmit antennas ($M=2$) and serves two single-antenna $\alpha$-users ($\mathcal{K}_{\alpha}=\{1,2\}$) and 60 single-antenna $0$-users ($\mathcal{K}_{0}=\{3,4,\ldots,62\}$). In the second setting, the BS has four transmit antennas ($M=4$) and serves  four single-antenna $\alpha$-users ($\mathcal{K}_{\alpha}=\{1,2,3,4\}$) and 60 single-antenna $0$-users ($\mathcal{K}_{0}=\{5,6,\ldots,64\}$). The CSIT quality factor of $\alpha$-users is fixed to $\alpha=0.5$.
Fig. \ref{fig:Plot_1 a} illustrates the ESR (i.e., $u_k=1,\forall k\in\mathcal{K}_{\alpha}$) of $\alpha$-users versus their long-term SNR for both TP--RSMA and PP--RSMA strategies when  $\alpha=0.5$ and $\theta=0.5$. To ensure the fairness of comparison, the ERs of all $0$-users are maintained to be the same for both TP--RSMA and PP--RSMA. This is achieved by tuning the power allocated to the data stream of the $0$-users  while using RS for $\alpha$-users in PP--RSMA. Two different cases are considered where the long-term SNR of the $0$-users is 10 dB or 20 dB lower than that of the $\alpha$-users. 
As the $\alpha$-and $0$-users are scheduled separately in TP--RSMA, the ESR remains the same in the specified two cases. The difference in SNR of the $0$-users only affects the ESR  in PP--RSMA. According to Fig. \ref{fig:Plot_1 a}, it is evident that the proposed PP--RSMA achieves obvious ESR gain over TP--RSMA in both cases. As the difference between the SNR of $\alpha$-users and that of $0$-users becomes larger, the ESR gain increases, which cannot be observed from DoF analysis.

\vspace{-1mm}
\section{Problem Formulation}
\vspace{-1mm}
\label{sec: problem formulation}
 The capacity region and  the corresponding capacity achieving strategy of overloaded multi-antenna networks with heterogeneous CSIT remains an open problem. We therefore explore the achievable rate regions achieved by TP--RSMA and PP--RSMA. The optimized precoders obtained by solving the ESR problem of PP--RSMA specified in this section  is guaranteed to achieve the optimal DoF performance characterized in Theorem \ref{theorem 1}.

Following \cite{RS2016hamdi}, we aim at  designing precoders at BS  to maximize the ESR of $\alpha$-users for the transmission carried out over a long sequence of channel uses and the problem is subject to the long-term QoS rate constraints of all users (including  $\alpha$-users and $0$-users) and  long-term  transmit power constraint. For a given  scheme  $\textrm{X}\in$\{``{TP}",``{PP}"\}, the corresponding ESR  is defined as $ \sum_{k\in\mathcal{K}_{\alpha}} \overline{R}_{k}^{\textrm{X}}$. 
However, such ESR problem requires to design precoders (including power) across different channel states,  which is intractable. To ensure the tractability of the problem, we follow  \cite{RS2016hamdi} to replace the long-term transmit power constraint  with short-term transmit power constraints for each channel state
and equivalently transform the ESR problem to the  Average Sum Rate (ASR) maximization problem in the following.

\vspace{-2mm}
\subsection{Average Sum Rate Maximization Problem}
We first  introduce the definition of Average Rates (ARs) as specified in Definition \ref{def: AR}. AR is defined as the expectation of rate over the CSIT uncertainty for a given instantaneous channel estimate. It is accessible at the BS since the long-term conditional density  $f_{\mathbf{H}\mid\widehat{\mathbf{H}}}(\mathbf{H}\mid\widehat{\mathbf{H}})$ is perfectly known at the BS. 
\begin{definition}
	\label{def: AR}
	The AR $\widehat{R}_{i,k}$ of decoding stream $s_{i},  i\in\{0,c,p\}$ at user $k,k\in\mathcal{K}$ for a given channel estimate $\widehat{\mathbf{H}}$ and precoder $\mathbf{P}(\widehat{\mathbf{H}})$  is\footnote{When $i=p$, the stream to be decoded at user $k$ is $s_p=s_k$. } 
	\vspace{-1mm}
	\begin{equation}
	\label{eq: average rate}
	\begin{aligned}
	\widehat{R}_{i,k}=\mathbb{E}_{\mathbf{H}\mid \widehat{\mathbf{H}}}\left\{\log\left(1+\gamma_{i,k}\right) \mid \widehat{\mathbf{H}}\right\}, 
	\end{aligned}
	\end{equation}
	where  $\gamma_{i,k}$ is specified in (\ref{eq: SINR cp}) when $i\in\{c,p\}$. $\gamma_{0,k}=\gamma_{0,k}^{\textrm{TP}}$ if we consider TP--RSMA and $\gamma_{0,k}=\gamma_{0,k}^{\textrm{PP}}$  if we consider PP--RSMA.   
\end{definition}

Notice that AR is different from ER since ER considers all possible channel states  while AR considers one channel  only. Their relation is obtained as \cite{RS2016hamdi}:
\begin{equation}
\overline{R}_{i,k}=\mathbb{E}_{ \widehat{\mathbf{H}}}\left\{\mathbb{E}_{\mathbf{H}\mid \widehat{\mathbf{H}}}\left\{\log\left(1+\gamma_{i,k}\right)\mid \widehat{\mathbf{H}}\right\}\right\}=\mathbb{E}_{ \widehat{\mathbf{H}}}\left\{\widehat{R}_{i,k}\right\}.
\end{equation}
Define the portion of  AR  corresponding to the rate of 
$W_{c,k}$ as $\widehat{C}_{k}$
such that 
$\sum_{k \in \mathcal{K}_{\alpha}} \widehat{C}_{k} = \widehat{R}_{c}$, where $\widehat{R}_{c}$ is the  AR of the common stream that must not exceed $ \min\{ \widehat{R}_{c,k} \mid k\in \mathcal{K}_{\alpha}\}$.   $\widehat{R}_{c}$  is obtained by approximating the ER of the common stream  $\overline{R}_{c}$ using its lower bound \cite{RS2016hamdi}, which is given as
\begin{equation}
\min_{k\in \mathcal{K}_{\alpha}}\left\{\mathbb{E}_{\widehat{\mathbf{H}}}\left\{	\widehat{R}_{c,k}\right\}\right\} \geq \mathbb{E}_{ \widehat{\mathbf{H}}}\left\{	\min\left\{ \widehat{R}_{c,k}\mid k\in \mathcal{K}_{\alpha}\right\}\right\}.
\end{equation}
Therefore, the respective AR of each user for TP--RSMA and PP--RSMA  are given as 
\begin{equation}
\label{eq: AR TP PP}
\begin{aligned}
\widehat{R}_{k}^{\textrm{TP}}  &=\left\{\begin{matrix}
\widehat{\theta} \big( \widehat{C}_{k} + \widehat{R}_{p,k} \big), 
& \forall k \in \mathcal{K}_{\alpha} \\ 
\theta_{0,k}   (1-\widehat{\theta}) \widehat{R}_{0,k},
&  \forall k \in 
\mathcal{K}_{0}
\end{matrix}\right., \,\,\,\,
\widehat{R}_{k}^{\textrm{PP}} =\left\{\begin{matrix}
\widehat{C}_{k} + \widehat{R}_{p,k}, 
& \forall k \in \mathcal{K}_{\alpha} \\ 
\theta_{0,k} \min_{i \in \mathcal{K}_{\alpha} \cup \{k\} } \{\widehat{R}_{0,k}\},
&  \forall k \in 
\mathcal{K}_{0}
\end{matrix}\right.,
\end{aligned}
\end{equation}
where $\widehat{\theta}$ is a short-term TP factor adaptive to each channel state.  Following the Law of Large Number (LLN), ER  of each user is  approximated by averaging AR  over all channel states, i.e., 
$
\overline{R}_{k}^{\textrm{X}} \approx\frac{1}{T}\sum_{t=1}^{T}  \widehat{R}_{k}^{\textrm{X}}(t)
$
if $T$ is a sufficiently large number, where
$\widehat{R}_{k}^{\textrm{X}}(t)$ is the AR for a given channel state $\mathbf{H}(t)$. 
With the  AR introduced in (\ref{eq: AR TP PP}), the long-term QoS ER constraints can be replaced with the short-term AR for each channel state, i.e., $\{\widehat{R}_{k}^{\textrm{X}}\geq R_k^{th}\}_{{\mathbf{H}}}$, where $\textrm{X}\in$\{``{TP}",``{PP}"\}.
The ESR objective function is  also approximated as
$
\sum_{k\in\mathcal{K}_{\alpha}} \overline{R}_{k}^{\textrm{X}} \approx\frac{1}{T}\sum_{t=1}^{T} 	\sum_{k\in\mathcal{K}_{\alpha}}\widehat{R}_{k}^{\textrm{X}}(t)
$.

Based on  the above approximation, the dependencies among channel states are removed. Maximizing the ESR under  short-term QoS average rate and transmit power constraints is achieved by optimizing precoders such that the ASR, i.e.
$\sum_{k\in\mathcal{K}_{\alpha}} \widehat{R}_{k}^{\textrm{X}}(t)$, is maximized for each
$\widehat{\mathbf{H}}(t)$. 
Focusing on an arbitrary channel use and dropping the time index $t$,  we aim at solving the respective ASR optimization problems for TP--RSMA and PP--RSMA as
\begin{subequations}
	\label{eq: opt ASR TP}
	\begin{align}
	&	\max_{\widehat{\theta}, \mathbf{{P}}, \widehat{\mathbf{c}}}  \sum_{k\in\mathcal{K}_{\alpha}} \widehat{R}_{k}^{\textrm{TP}}\\
	\mbox{s.t.}\,\,
	&  \,\, \widehat{R}_{k}^{\textrm{TP}}\geq R_k^{th}, \forall k\in\mathcal{K} \\
	&  \,\,\sum_{k' \in \mathcal{K}_{\alpha}} \widehat{C}_{k'} \leq  \widehat{R}_{c,k},  \forall k\in \mathcal{K}_{\alpha} \label{c: sum common AR} \\
	&  \,\,	\mathrm{tr}(\mathbf{P}_{\alpha}\mathbf{P}_{\alpha}^{\Hrm}) \leq P \label{c: pt1 TP}\\
	&\,\,\| \mathbf{p}_{0} \|^{2} \leq P \label{c: pt2 TP}\\
	&\,\,	\mathbf{\widehat{c}}\geq \mathbf{0}  \label{c: common AR vector} \\
	&\,\, 0\leq\widehat{\theta}\leq 1
	\end{align}
\end{subequations}
and
\begin{subequations}
	\label{eq: opt ASR PP}
	\begin{align}
	&	\max_{ \mathbf{{P}}, \widehat{\mathbf{c}}}  \sum_{k\in\mathcal{K}_{\alpha}} \widehat{R}_{k}^{\textrm{PP}}\\
	\mbox{s.t.}\,\,
	&  \,\, \widehat{R}_{k}^{\textrm{PP}}\geq R_k^{th}, \forall k\in\mathcal{K} \\
	&  \,\,	\mathrm{tr}(\mathbf{P}_{\alpha}\mathbf{P}_{\alpha}^{\Hrm})+\| \mathbf{p}_{0} \|^{2} \leq P \label{c: pt PP}\\
	&\,\,	\textrm{(\ref{c: sum common AR})}, \textrm{(\ref{c: common AR vector})} 
	\end{align}
\end{subequations}
where  $	\widehat{\mathbf{c}} =[\widehat{C}_{1} \; \widehat{C}_{2} \;\cdots \;\widehat{C}_{M}]$  is the AR allocation for the common stream $s_c$.  It is required to be jointly optimized with the precoder so as to maximize the ASR.  $R_k^{th}$ is the QoS rate requirement of user $k$. 
The TP factor $\widehat{\theta}$ is jointly optimized with precoders in order to maximize ASR. In PP--RSMA, constraint  (\ref{c: pt PP}) is a  sum power constraint for all precoders since all users in $\mathcal{K}$ are served simultaneously. This contrasts with separated power constraints (\ref{c: pt1 TP}) and (\ref{c: pt2 TP}) for TP--RSMA  where  users in $\mathcal{K}_{\alpha}$ and  $\mathcal{K}_0$ are served in orthogonal time phases.

\vspace{-1.5mm}
\section{Optimization Framework} \label{sec: optimization framework}
In this section, the optimization frameworks proposed to solve the stochastic optimization problems (\ref{eq: opt ASR TP}) and (\ref{eq: opt ASR PP}) are illustrated.  Specifically, we follow the  approach proposed in \cite{RS2016hamdi}  to  transform the original stochastic problems into the deterministic problems by using the Sample Average Approximation (SAA)  approach and the transformed problems are solved by using the Weighted Minimum Mean Square Error (WMMSE)-based optimization algorithms. In the following, we first specify the SAA and WMMSE-based algorithm to solve PP--RSMA problem (\ref{eq: opt ASR PP}) and then  extend it to solve TP--RSMA problem   (\ref{eq: opt ASR TP}). 
\vspace{-0.5cm}
\subsection{Sample Average Approximation (SAA)  approach}
\vspace{-0.1cm}
Though the BS only has partial CSI of users in $\mathcal{K}_{\alpha}$ and no knowledge of CSI of users in $\mathcal{K}_{0}$, a channel sample of size $N$ (indexed by $\mathcal{N}=\{1, \ldots,N\}$) can be generated since the conditional density $f_{{\mathbf{H}}\mid \widehat{\mathbf{H}}}({\mathbf{H}}\mid \widehat{\mathbf{H}})$ is  known at the BS, which is given by
\begin{equation}
\mathbb{H}^{(N)}\triangleq \left\{\mathbf{H}^{(n)}=\widehat{\mathbf{H}}+\widetilde{\mathbf{H}}^{(n)}\mid \widehat{\mathbf{H}}, n\in\mathcal{N}\right\},
\end{equation}
where $\widehat{\mathbf{H}}=[\widehat{\mathbf{h}}_1\cdots \widehat{\mathbf{h}}_{M} \;\mathbf{0}]$ with $\mathbf{0}\in\mathbb{C}^{M\times (K-M)}$ for $0$-users.
Under the assumption of finite SNR and bounded transmit power, the AR $\widehat{R}_{i,k}$ specified in equation (\ref{eq: average rate}) is approximated  as
\vspace{-2mm}
\begin{equation}
\label{eq: SAF}
\widehat{R}_{i,k}\approx \underset{{\widehat{R}_{i,k}^{(N)}}(\widehat{\mathbf{H}})}{{\underbrace{\frac{1}{N}\sum_{n=1}^{N}\log\left(1+\gamma_{i,k}^{(n)}\right)}}},
\vspace{-2mm}
\end{equation}
where  $\gamma_{i,k}^{(n)}$ is the SINR calculated for the  $n$th channel sample  in $\mathbb{H}^{(N)}$. The precoder $\mathbf{P}$ in (\ref{eq: SAF}) remains static over all the $N$ channel samples. $\widehat{R}_{i,k}^{(N)}(\widehat{\mathbf{H}})$ is the approximated AR. Based on 
 LLN, we have $\widehat{R}_{i,k}=\lim_{N\to\infty}\widehat{R}_{i,k}^{(N)}(\widehat{\mathbf{H}})$.

By adopting the aforedescribed SAA approach, the stochastic optimization problem (\ref{eq: opt ASR PP}) is approximated by the following deterministic problem with the average common and private rates approximated by equation (\ref{eq: SAF}) for all  channel samples, which is given by
\begin{subequations}
	\label{eq: opt SAA PP}
	\begin{align}
	&	\max_{ \mathbf{{P}}, \widehat{\mathbf{c}}}  \sum_{k\in\mathcal{K}_{\alpha}}\widehat{C}_{k} + \widehat{R}_{p,k}^{(N)}\\
	\mbox{s.t.}\,\,
	&  \,\, \widehat{C}_{k} + \widehat{R}_{p,k}^{(N)}\geq R_k^{th}, \forall k\in\mathcal{K}_{\alpha} \\
	&\theta_{0,k} \min_{i \in \mathcal{K}_{\alpha} \cup \{k\} } \{\widehat{R}_{0,i}^{(N)}\}\geq R_k^{th},  \forall k \in 	\mathcal{K}_{0}\\
	&  \,\,\sum_{k' \in \mathcal{K}_{\alpha}} \widehat{C}_{k'} \leq  \widehat{R}_{c,k}^{(N)}, \forall k\in\mathcal{K}_{\alpha}\\
	&\,\,	\textrm{(\ref{c: common AR vector})}, \textrm{(\ref{c: pt PP})}
	\end{align}
\end{subequations}
\vspace{-0.9cm}
\subsection{WMMSE Algorithm}
Though problem (\ref{eq: opt SAA PP}) is still non-convex, it fits to be solved by WMMSE algorithm which is suited for  non-convex optimization problems when non-convexity comes from rate expressions. The WMMSE algorithm proposed in \cite{RS2016hamdi} is extended to solve  (\ref{eq: opt SAA PP}).

All users first employ respective equalizers $\{g_{0,k}\mid k\in\mathcal{K}\}$ to decode $s_0$, the estimated common stream for $0$-user is $\widehat{s}_{0,k}=g_{0,k}y_k$. For $\alpha$-users, once $s_0$ is decoded and  removed from the received signal, the equalizers  $\{g_{c,k}\mid k\in\mathcal{K}_{\alpha}\}$ are used at the corresponding $\alpha$-users to decode $s_c$ followed by  the equalizers  $\{g_{k}\mid k\in\mathcal{K}_{\alpha}\}$ to decode private streams. The estimated common and private streams at each $\alpha$-user $k$ are given by $\widehat{s}_{c,k}=g_{c,k}(y_k-\mathbf{h}_k^{\Hrm}\mathbf{p}_0\widehat{s}_{0,k})$ and $\widehat{s}_{p,k}=g_{k}(y_k-\mathbf{h}_k^{\Hrm}\mathbf{p}_0\widehat{s}_{0,k}-\mathbf{h}_k^{\Hrm}\mathbf{p}_c\widehat{s}_{c,k})$. 
The Mean Square Error (MSE) of decoding the corresponding stream $s_i$ at user $k$ is defined as
\begin{equation}
\label{eq:MSE}
\begin{aligned}
&\varepsilon_{i,k}=\mathbb{E}\{|\widehat{s}_{i,k}-s_{i}|^{2}\}=|g_{k}^i|^2T_{i,k}-2\Re\{g_{k}^i\mathbf{h}_{k}^{\Hrm}\mathbf{p}_{i}\}+1,
\end{aligned}
\end{equation}
where $i\in\{0,c,p\}$ if $k\in\mathcal{K}_{\alpha}$ and $i=0$ if $k\in\mathcal{K}_{0}$. $T_{p,k}=\sum_{j\in\mathcal{K}_{\alpha}}|\mathbf{h}_{k}^{\Hrm}\mathbf{p}_{j}|^2+1$, $T_{c,k}=T_{p,k}+|\mathbf{h}_{k}^{\Hrm}\mathbf{p}_{c}|^2$ and $T_{0,k}=T_{c,k}+|\mathbf{h}_{k}^{\Hrm}\mathbf{p}_{0}|^2$. 

The Weighted MSE (WMSE) of decoding $s_i$ at user $k$ is defined as 
\vspace{-1.5mm}
\begin{equation}
	\xi_{i,k}= w_{i,k}\varepsilon_{i,k}-\log_{2}(w_{i,k}),
\vspace{-1.5mm}
\end{equation}
where $w_{i,k}$ is the weight associated with MSE $\varepsilon_{i,k}$.
By minimizing $\xi_{i,k}$ over $w_{i,k}$ and $g_{i,k}$,  the corresponding WMMSE metric is defined as
\begin{equation}
\xi_{i,k}^{\textrm{MMSE}}=\min_{w_{i,k},g_{i,k}}\xi_{i,k}.
\end{equation}

The following proposition is then obtained.
\begin{proposition}
	\label{prop 1}
	Define the  instantaneous rate of decoding stream $s_i$  at user $k$ as $R_{i,k}=\log\left(1+\gamma_{i,k}\right)$, the relation between $R_{i,k}$ and its WMMSE  follows that:
		\vspace{-1.5mm}
	\begin{equation}
	\label{eq: rate-wmmse}
	\xi_{i,k}^{\textrm{MMSE}}=1-R_{i,k}. 
	\vspace{-1.5mm}
	\end{equation}
\end{proposition}
\textit{Proof:} Readers are referred to \cite{RS2016hamdi} for more details of proof. To briefly summarize, the equivalence in (\ref{eq: rate-wmmse}) is due to the fact that the optimal equalizer calculated by 
$\frac{\partial\xi_{i,k}}{\partial g_{i,k}}=0$ is
\vspace{-1.5mm}
\begin{equation}
\label{eq: WMMSE equalizers}
g_{i,k}^{\textrm{MMSE}}=\mathbf{p}_{i}^{\Hrm}\mathbf{h}_{k}({T}_{i,k})^{-1}.
\vspace{-1.5mm}
\end{equation} 
By further solving  $\left.\frac{\partial\xi_{i,k}}{\partial w_{i,k}}\right|_{g_{i,k}=g_{i,k}^{\textrm{MMSE}}}=0$, we have
\vspace{-0.5mm}
\begin{equation}
\label{eq: WMMSE weights}
w_{i,k}^{\textrm{MMSE}}={\varepsilon_{i,k}^{-1}(g_{i,k}^{\textrm{MMSE}})}=(T_{i,k}-|\mathbf{h}_{k}^{\Hrm}\mathbf{p}_{i}|^2)^{-1}T_{i,k}.
\vspace{-1mm}
\end{equation} 
Substituting ($w_{i,k}^{\textrm{MMSE}},g_{i,k}^{\textrm{MMSE}}$) back to $\xi_{i,k}$,  we obtain that  $\xi_{i,k}^{\textrm{MMSE}}=\log_2(w_{i,k}^{\textrm{MMSE}})=1-R_{i,k}$. The proof is completed. \qed

Proposition \ref{prop 1} establishes the connection between rate and WMMSE metric for a given instantaneous channel use. It is further extended to the AR-WMMSE relationship by approximating the AR using average WMMSE over a channel sample $\mathbb{H}^{(N)}$ of size $N$. Define the Average WMSE (AWMSE) of decoding stream $s_i$ at user $k$ as
\begin{equation}
\widehat{\xi}_{i,k}^{(N)}={\frac{1}{N}\sum_{n=1}^{N}\xi_{i,k}^{(n)}}, 
\end{equation} 
where $\xi_{i,k}^{(n)}$ is the WMSE calculated for the  $n$th channel sample  in $\mathbb{H}^{(N)}$ and $w_{i,k}^{(n)},g_{i,k}^{(n)}$ are associated with the $n$th channel sample in $\mathbb{H}^{(N)}$.  For compactness, we also define  $\mathbf{w}_{i,k}=\{w_{i,k}^{(n)} \mid n\in\mathcal{N}\}$  and $\mathbf{g}_{i,k}=\{g_{i,k}^{(n)} \mid n\in\mathcal{N}\}$ as the respective MSE weights and equalizers for user $k$ to decode $s_i$. We obtain the following AR--WMMSE relationship 
\begin{equation}
\label{eq: rate-wmmse average}
\widehat{\xi}_{i,k}^{\textrm{MMSE}^{(N)}}=\min_{\mathbf{w}_{i,k},\mathbf{g}_{i,k}}\widehat{\xi}_{i,k}^{(N)}=1-{\widehat{R}_{i,k}^{(N)}},
\end{equation}
which can be proved by employing the same method as in Proposition \ref{prop 1}.

Based on (\ref{eq: rate-wmmse average}),  the target problem (\ref{eq: opt SAA PP})  for PP--RSMA is then equivalently transformed   into the following WMMSE problem:
\begin{subequations}
	\label{eq: RS WMMSE}
	\begin{align}
	&	\min_{\mathbf{{P}},\widehat{\mathbf{x}},\mathbf{W},\mathbf{G}}\,\, \sum_{k\in\mathcal{K}_{\alpha}}\widehat{X}_{k}+\widehat{\xi}_{p,k}^{{(N)}}\\
	\mbox{s.t.}\,\,
	&\,\, \widehat{X}_{k}+\widehat{\xi}_{p,k}^{{(N)}} \leq 1- R_{k}^{th}, \forall k\in\mathcal{K}_{\alpha} \label{c1_RS WMMSE}\\
		&\theta_{0,k} \max_{i \in \mathcal{K}_{\alpha} \cup \{k\} } \{\widehat{\xi}_{0,i}^{{(N)}}\}\leq \theta_{0,k}-R_k^{th},  \forall k \in 	\mathcal{K}_{0}\\
	&  \,\,\sum_{k' \in \mathcal{K}_{\alpha}} \widehat{X}_{k'}+1 \geq  \widehat{\xi}_{c,k}^{(N)}, \forall k\in\mathcal{K}_{\alpha}\\
	&\,\,	\widehat{\mathbf{x}} \leq \mathbf{0}  \label{c: x vector} \\
	&\,\,\textrm{(\ref{c: pt PP})},
	\end{align}
\end{subequations}
where $\widehat{\mathbf{x}}=[\widehat{X}_{1} \; \widehat{X}_{2} \;\cdots \;\widehat{X}_{M}]$ is the transformation of $\widehat{\mathbf{c}}$. It follows that $\widehat{\mathbf{x}}=-\widehat{\mathbf{c}}$.   $\mathbf{W}\triangleq\{\mathbf{w}_{0,k'},\mathbf{w}_{c,k},\mathbf{w}_{p,k} \mid k\in\mathcal{K}_{\alpha}, k'\in\mathcal{K}\}$  and $\mathbf{G}\triangleq\{\mathbf{g}_{0,k'},\mathbf{g}_{c,k},\mathbf{g}_{p,k} \mid k\in\mathcal{K}_{\alpha}, k'\in\mathcal{K}\}$ are the respective MSE weights and equalizers.

Though the joint optimization of variables $(\mathbf{{P}},\widehat{\mathbf{x}},\mathbf{W},\mathbf{G})$ in (\ref{eq: RS WMMSE})  is still non-convex,  it has the property of block-wise convexity in each block of $(\mathbf{{P}},\widehat{\mathbf{x}})$, $\mathbf{W}$ and $\mathbf{G}$ if other two blocks are fixed, which fits to be solved by using Alternating Optimization (AO) as illustrated in Algorithm \ref{WMMSE algorithm}. At the $[t]$-th iteration, the respective optimal solutions of equalizers and weights are $\mathbf{G}^{[t]}=\mathbf{G}^{\textrm{MMSE}}(\mathbf{P}^{[t-1]})$, $\mathbf{W}^{[t]}=\mathbf{W}^{\textrm{MMSE}}(\mathbf{P}^{[t-1]})$ with each corresponding element calculated by (\ref{eq: WMMSE equalizers}) and (\ref{eq: WMMSE weights}). Then, we obtain the problem to update $(\mathbf{{P}}^{[t]},\widehat{\mathbf{x}}^{[t]})$ at the $[t]$-th iteration by substituting $\mathbf{G}^{[t]},\mathbf{W}^{[t]}$ back to (\ref{eq: RS WMMSE}), which is given by
\begin{equation}
	\label{eq: RS WMMSE final}
	\resizebox{.93 \textwidth}{!} {$
	\begin{aligned}
	&	\min_{\mathbf{{P}},\widehat{\mathbf{x}}}\,\, \sum_{k\in\mathcal{K}_{\alpha}}(\widehat{X}_{k}+\sum_{{j\in\mathcal{K}_{\alpha}}}\mathbf{p}_{j}^{\Hrm}\bar{\Psi}_{p,k}\mathbf{p}_{ j}+\bar{t}_{p,k}-2\Re\left\{\bar{\mathbf{f}}_{p,k}^{\Hrm}\mathbf{p}_{k}\right\}+\bar{w}_{p,k}-\bar{\nu}_{p,k})\\
	\mbox{s.t.}\,\,
	&\,\, \widehat{X}_{k}+\sum_{{j\in\mathcal{K}_{\alpha}}}\mathbf{p}_{j}^{\Hrm}\bar{\Psi}_{p,k}\mathbf{p}_{ j}+\bar{t}_{p,k}-2\Re\left\{\bar{\mathbf{f}}_{p,k}^{\Hrm}\mathbf{p}_{k}\right\}+\bar{w}_{p,k}-\bar{\nu}_{p,k}\leq 1- R_{k}^{th}, \forall k\in\mathcal{K}_{\alpha}\\
	&\theta_{0,k} \max_{i \in \mathcal{K}_{\alpha} \cup \{k\} } \left\{\sum_{{j\in\mathcal{K}_{\alpha} \cup \{c,0\} }}\mathbf{p}_{j}^{\Hrm}\bar{\Psi}_{0,i}\mathbf{p}_{ j}+\bar{t}_{0,i}-2\Re\left\{\bar{\mathbf{f}}_{0,i}^{\Hrm}\mathbf{p}_{0}\right\}+\bar{w}_{0,i}-\bar{\nu}_{0,i}\right\}\leq \theta_{0,k}-R_k^{th},  \forall k \in 	\mathcal{K}_{0}\\
	&  \,\,\sum_{k' \in \mathcal{K}_{\alpha}} \widehat{X}_{k'}+1 \geq  \sum_{{j\in\mathcal{K}_{\alpha} \cup \{c\} }}\mathbf{p}_{j}^{\Hrm}\bar{\Psi}_{c,k}\mathbf{p}_{ j}+\bar{t}_{c,k}-2\Re\left\{\bar{\mathbf{f}}_{c,k}^{\Hrm}\mathbf{p}_{c}\right\}+\bar{w}_{c,k}-\bar{\nu}_{c,k}, \forall k\in\mathcal{K}_{\alpha}\\
	&\,\,\textrm{(\ref{c: pt PP})}, \textrm{(\ref{c: x vector})} 
	\end{aligned}$}
\end{equation}
where  
$\bar{\Psi}_{i,k},  \bar{t}_{i,k},  \bar{\mathbf{f}}_{i,k}, \bar{w}_{i,k},  \bar{\nu}_{i,k}, i\in\{0,c,p\}$ are averaged over $N$ channel samples, i.e., $\bar{w}_{i,k}=\frac{1}{N}\sum_{n=1}^N{w}_{i,k}^{(n)}$. ${\Psi}_{i,k}^{(n)},  {t}_{i,k}^{(n)},  {\mathbf{f}}_{i,k}^{(n)}, {\nu}_{i,k}^{(n)}$  are updated in $n$th channel sample  as
\vspace{-3mm}
\begin{equation}
\resizebox{.9\textwidth}{!} {$
 {t}_{i,k}^{(n)}=w_{i,k}^{(n)}\left|g_{i,k}^{(n)}\right|^2, \,\,
  {\Psi}_{i,k}^{(n)}={t}_{i,k}^{(n)}\mathbf{h}_{k}^{(n)}(\mathbf{h}_{k}^{(n)})^{\Hrm},\,\,
\mathbf{f}_{i,k}^{(n)}=w_{i,k}^{(n)}\mathbf{h}_{k}^{(n)}(g_{i,k}^{(n)})^{\Hrm},\,\,
 {\nu}_{i,k}^{(n)}=\log_2\left(w_{i,k}^{(n)}\right).
$}
\vspace{-2mm}
\end{equation}
$(\mathbf{{P}}^{[t]},\widehat{\mathbf{x}}^{[t]})$ are updated by solving the convex Quadratically Constrained Quadratic Program (QCQP) (\ref{eq: RS WMMSE final}) via interior-point methods. The blocks $\mathbf{W}$,  $\mathbf{G}$ and $(\mathbf{P},\widehat{\mathbf{x}})$ are updated iteratively until the convergence of the ESR\footnote{The AO algorithm is guaranteed to converge to a stationary point of (\ref{eq: opt ASR PP}) given a feasible initial point. The proof of convergence is inline with \cite{RS2016hamdi}, which is not specified here  due to the space limitation. Readers are referred to \cite{RS2016hamdi} for more details.}.

\setlength{\textfloatsep}{5pt}	
\begin{algorithm}[t!]
	\textbf{Initialize}: $t\leftarrow0$, $\mathbf{P}$, $\mathrm{ASR}^{[t]}$\;
	\Repeat{$|\mathrm{ASR}^{[t]}-\mathrm{ASR}^{[t-1]}|\leq \epsilon$}{
		$t\leftarrow t+1$,	$\mathbf{P}^{[t-1]}\leftarrow \mathbf{P}$\;
		$\mathbf{W}^{[t]}\leftarrow\mathbf{W}^{\mathrm{MMSE}}(\mathbf{P}^{[t-1]})$; $\mathbf{W}^{[t]}\leftarrow\mathbf{G}^{\mathrm{MMSE}}(\mathbf{P}^{[t-1]})$\;
		update $(\mathbf{P},\widehat{\mathbf{x}})$ by solving (\ref{eq: RS WMMSE final});	
	}	
	\caption{WMMSE-based AO algorithm}
	\label{WMMSE algorithm}				
\end{algorithm}

When solving  problem (\ref{eq: opt ASR TP}) for TP--RSMA, though the TP factor $\widehat{\theta}$ increases the complexity to solve the problem, we observe that for a fixed $\widehat{\theta}$, the original problem  can be decomposed into  subproblems for $\alpha$-users and $0$-users. The subproblem for $\alpha$-users is the same as the one studied in \cite{RS2016hamdi} with an additional  QoS rate constraint for each user which can be solved directly by using the SAA and WMMSE algorithm proposed in \cite{RS2016hamdi} while the subproblem for $0$-users can be written into $\max_{\mathbf{p}_0} \min_{k\in\mathcal{K}_0}\widehat{R}_{k}^{\textrm{TP}}$ such that  constraints $\widehat{R}_{k}^{\textrm{TP}}\geq R_k^{th}, \forall k\in\mathcal{K}_0$ and $\| \mathbf{p}_{0} \|^{2} \leq P$  are met. After using SAA, it is a convex optimization problem which can be solved directly by using interior-point method. Therefore, by tuning $\widehat{\theta}$ and calculate the ASR for each $\widehat{\theta}$, the optimal $\widehat{\theta}$ to maximize the ASR in (\ref{eq: opt ASR TP}) can be obtained. 

\vspace{-3mm}
\section{Numerical Results} \label{sec: numerical results}
In this section, the ESR performance of the proposed TP--RSMA and PP--RSMA achieved by solving problem (\ref{eq: opt ASR TP}) and (\ref{eq: opt ASR PP}) is studied. Random channel generation is considered where the exact channel $\mathbf{h}_k$ of user $k, k\in\mathcal{K}$  has i.i.d. complex Gaussian entries drawn from the distribution $\mathcal{CN}(0,1)$ and the channel estimation error $\widetilde{\mathbf{h}}_k$ of $\alpha$-user $k, k\in\mathcal{K}_{\alpha}$  has entries drawn from distribution $\mathcal{CN}(0,\sigma_{e,k}^2)$ with $\sigma_{e,k}^2 = P^{-\alpha}$. The instantaneous CSIT of $0$-users is unknown.
The optimization frameworks specified in Section \ref{sec: optimization framework} is adopted to design the precoders for TP--RSMA and PP--RSMA in this subsection. We use the CVX toolbox \cite{grant2008cvx} to solve the approximated WMMSE problems. A transmission over $T=100$ channel users  is considered, i.e., $\mathcal{T}=\{1,\ldots,100\}$. For each channel use, the AR of each user is approximated using SAA method over the channel sample of size $N=1000$.
The precoders are initialized at the transmitter by the Maximum Ratio Transmission (MRT) and Singular Value Decomposition (SVD) method as studied in \cite{RS2016hamdi}. For both TP--RSMA and PP--RSMA, $\mathbf{p}_0$ is initialized by a random precoding vector with $p_0=||\mathbf{p}_0||^2$.     $\mathbf{p}_c$  is initialized as $\mathbf{p}_c=\sqrt{p_c}\widehat{\mathbf{p}}_c$ and $\widehat{\mathbf{p}}_c$ is the largest left singular vector of the channel estimate $[\widehat{\mathbf{h}}_1,\cdots,\widehat{\mathbf{h}}_M]$. The precoder $\mathbf{p}_k$ of the private stream  for $\alpha$-user $k$ is initialized as $\mathbf{p}_k=\sqrt{p_k}\frac{\widehat{\mathbf{h}}_k}{||\widehat{\mathbf{h}}_k||}$. $p_0,p_c,p_k$ are the power allocated to the corresponding precoders and they satisfy that $p_0+p_c+\sum_{k\in\mathcal{K}_{\alpha}}p_k=P$ if PP--RSMA is considered or $p_0+p_c=P, \sum_{k\in\mathcal{K}_{\alpha}}p_k=P$ if TP--RSMA is considered. 

\begin{figure}[t!]
	\centering
	\begin{subfigure}[b]{0.24\textwidth}
		\vspace{-1mm}
		\centering
		\includegraphics[width=\textwidth]{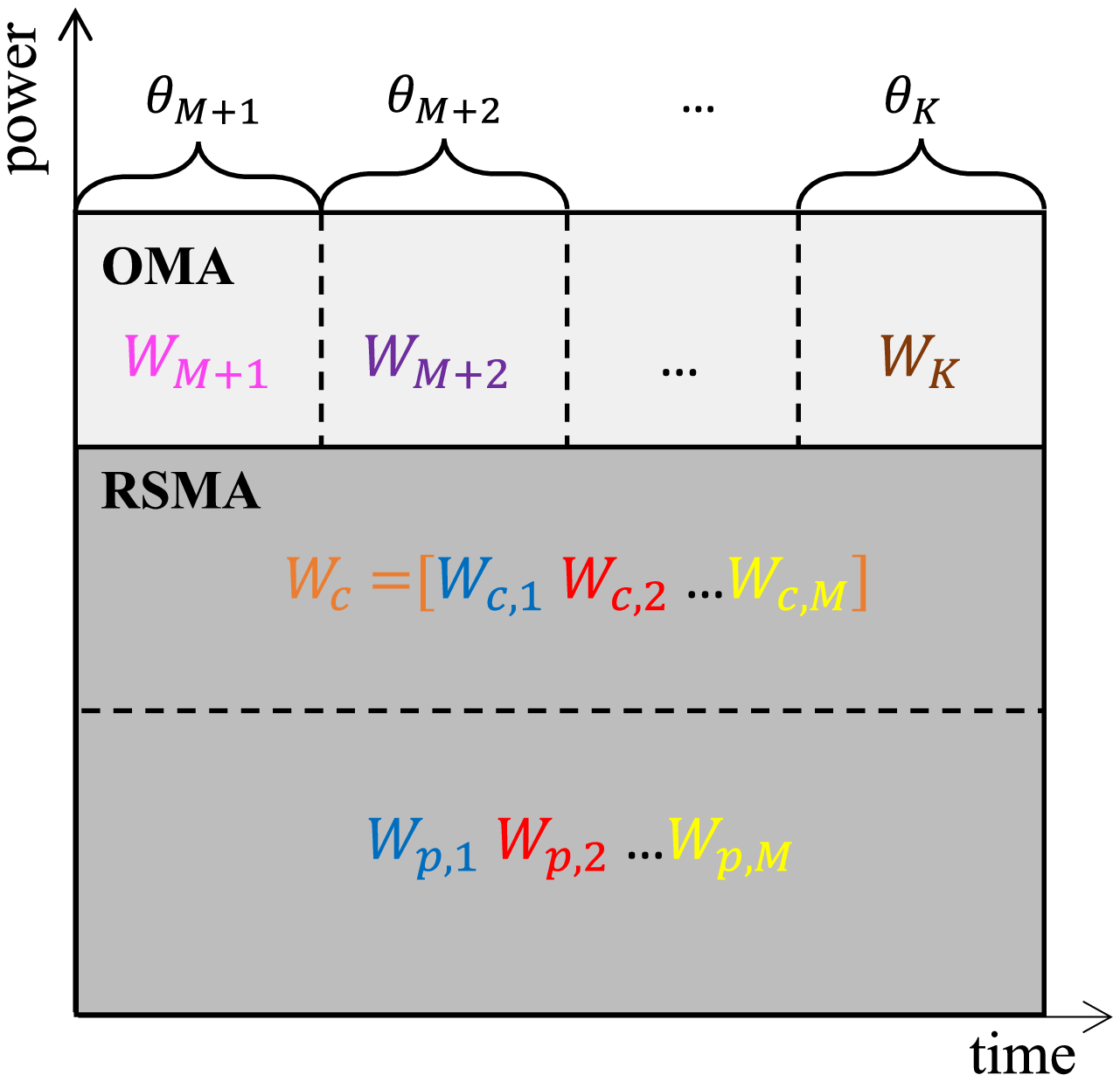}%
		\vspace{-1mm}
		\caption{PP--RSMA}
	\end{subfigure}%
	~
	\begin{subfigure}[b]{0.24\textwidth}
		\vspace{-1mm}
		\centering
		\includegraphics[width=\textwidth]{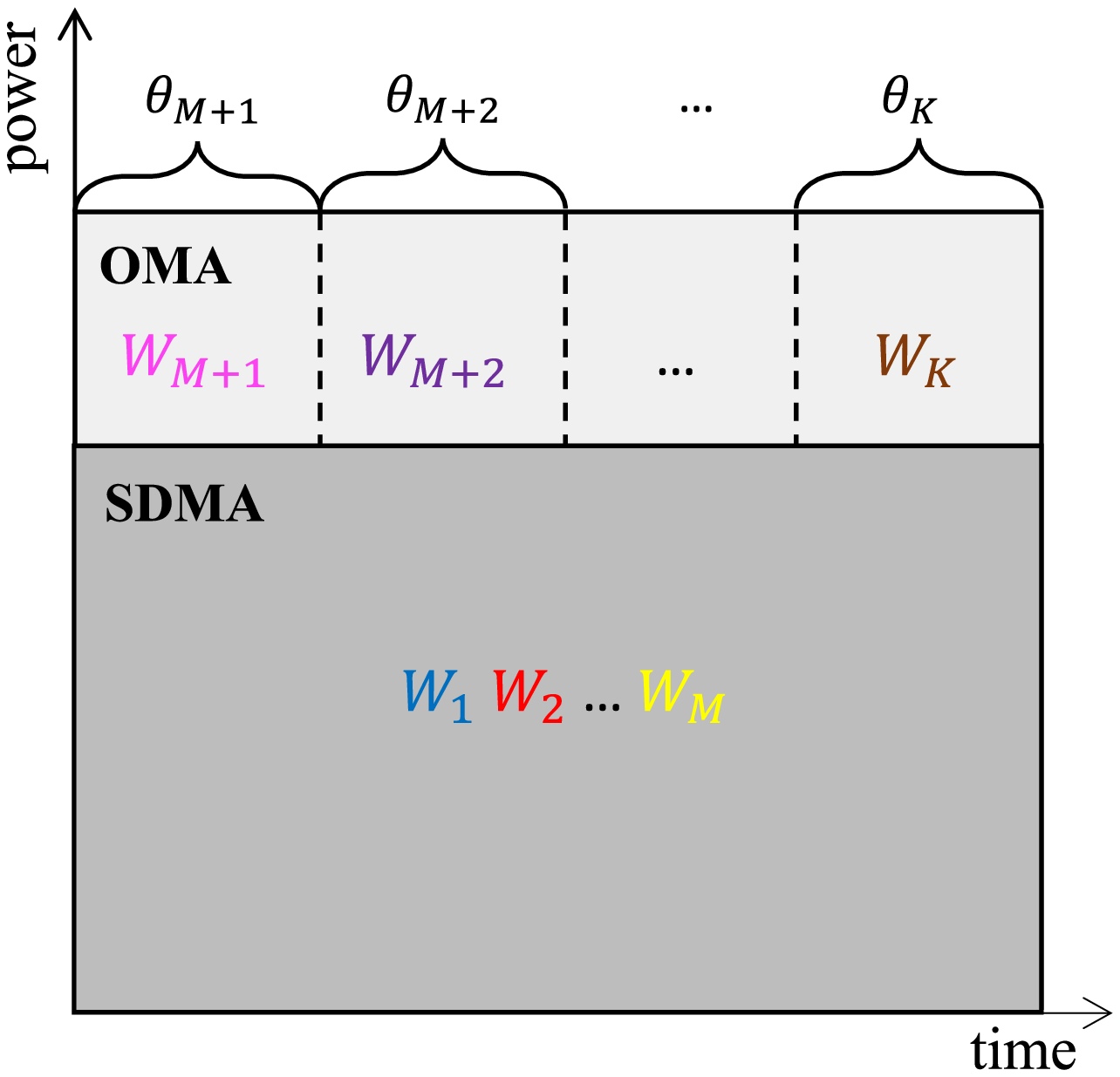}%
		\vspace{-1mm}
		\caption{PP--SDMA}
	\end{subfigure}%
	~
	\begin{subfigure}[b]{0.24\textwidth}
		\centering
		\includegraphics[width=\textwidth]{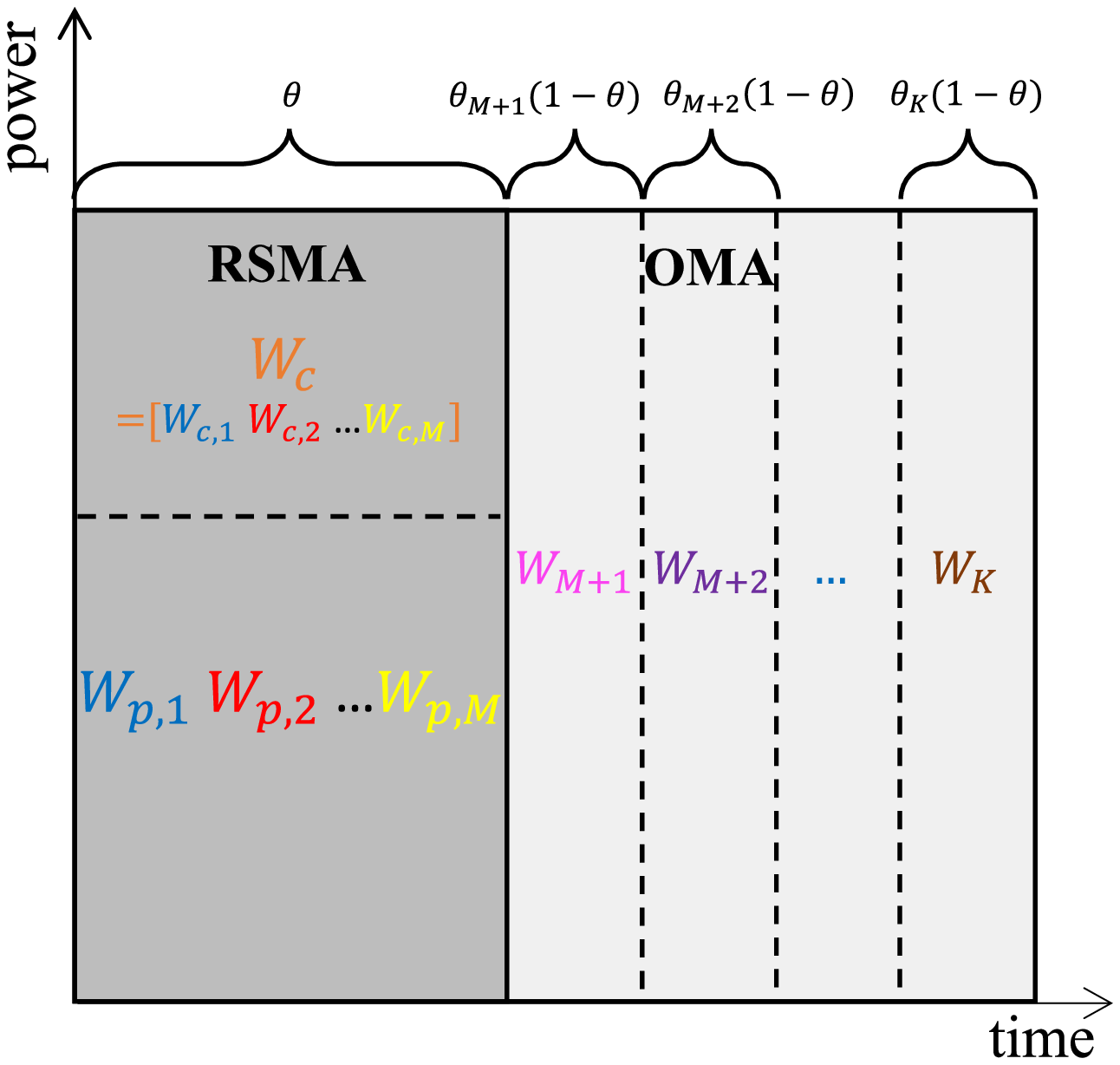}%
		\vspace{-1mm}
		\caption{TP--RSMA}
	\end{subfigure}%
	~
	\begin{subfigure}[b]{0.24\textwidth}
		\centering
		\includegraphics[width=\textwidth]{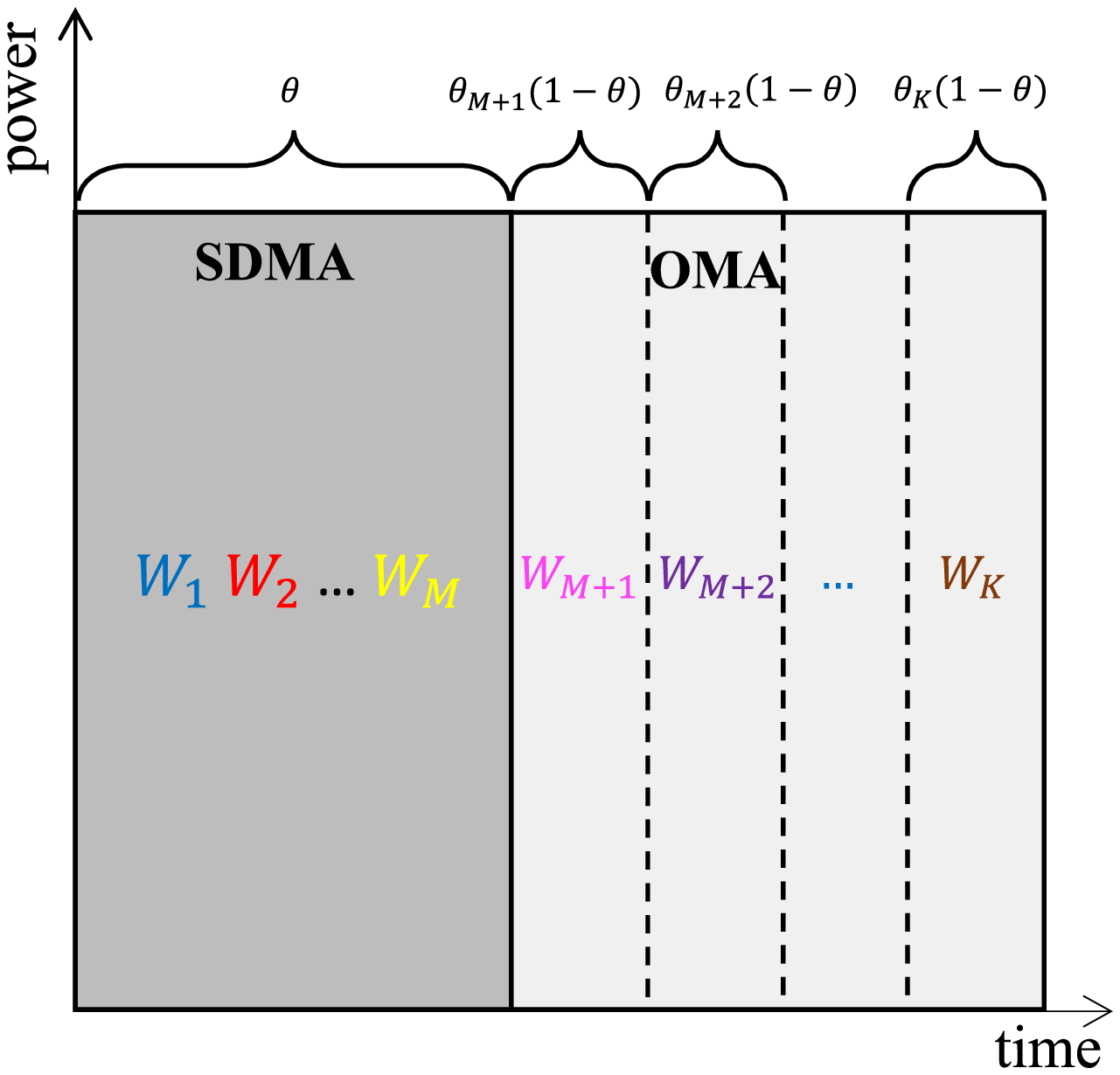}%
		\vspace{-1mm}
		\caption{TP--SDMA}
	\end{subfigure}%
	\vspace{-1mm}
	\caption{Comparison of the four transmission strategies in the numerical results}
	\label{fig: schemeComparison}
\end{figure}
The following four strategies as illustrated in Fig. \ref{fig: schemeComparison} are compared in the numerical results: 
\begin{itemize}
	\item \textbf{PP--RSMA}: the transmission approach we proposed in Section \ref{sec: PPRS}.
	\item \textbf{PP--SDMA}: users in $\mathcal{K}_{\alpha}$ and $\mathcal{K}_{0}$ are served via PP as in PP--RSMA. However, users in $\mathcal{K}_{\alpha}$ are served by SDMA as specified in \cite{mao2017rate}. After decoding and removing  $s_0$ from the received signal, user $k, k\in\mathcal{K}_{\alpha}$ directly decodes its intended stream.
	\item \textbf{TP--RSMA}: the transmission approach we proposed in Section \ref{sec: TPRS}.
	\item \textbf{TP--SDMA}: the $\alpha$-and $0$-users are served via TP and SDMA is adopted for the transmission of $\alpha$-users. 
\end{itemize}
As we mentioned previously, we assume uniform time allocation among $0$-users, i.e.,  $\theta_{0,k}=\frac{1}{K-M}$ for all strategies.
In the sequel, the ESR performance of  the above four strategies is evaluated in a wide range of user deployments considering a wide range of network load, CSIT qualities of $\alpha$-users and QoS rate requirements of all users.


\begin{figure}[t!]
	\vspace{-1mm}
	\centering
	 \begin{subfigure}[b]{0.33\textwidth}
		\vspace{-1mm}
		\centering
		\includegraphics[width=0.88\textwidth]{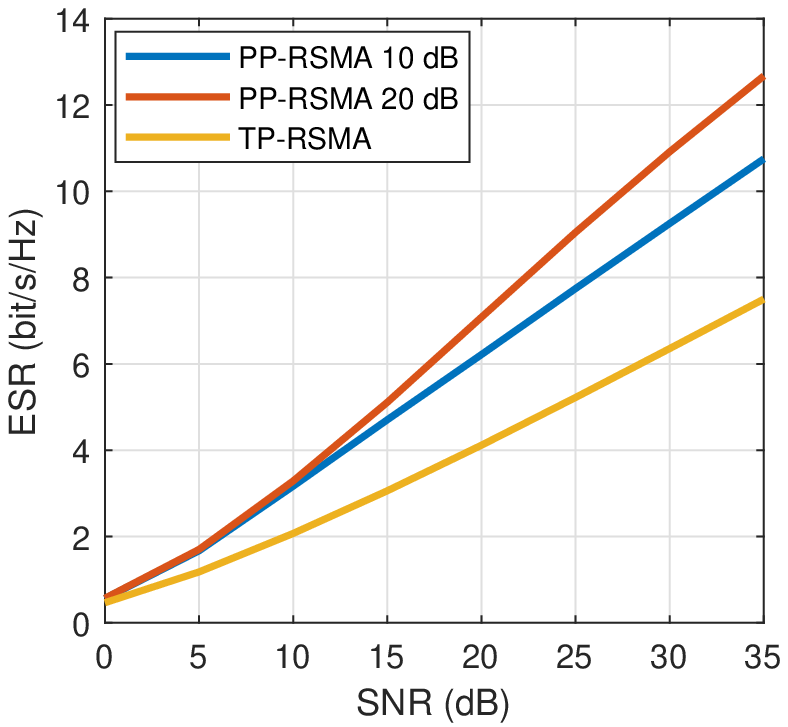}%
		\caption{Optimized precoding}
	\end{subfigure}%
	~
	\begin{subfigure}[b]{0.33\textwidth}
		\vspace{-1mm}
		\centering
		\includegraphics[width=0.88\textwidth]{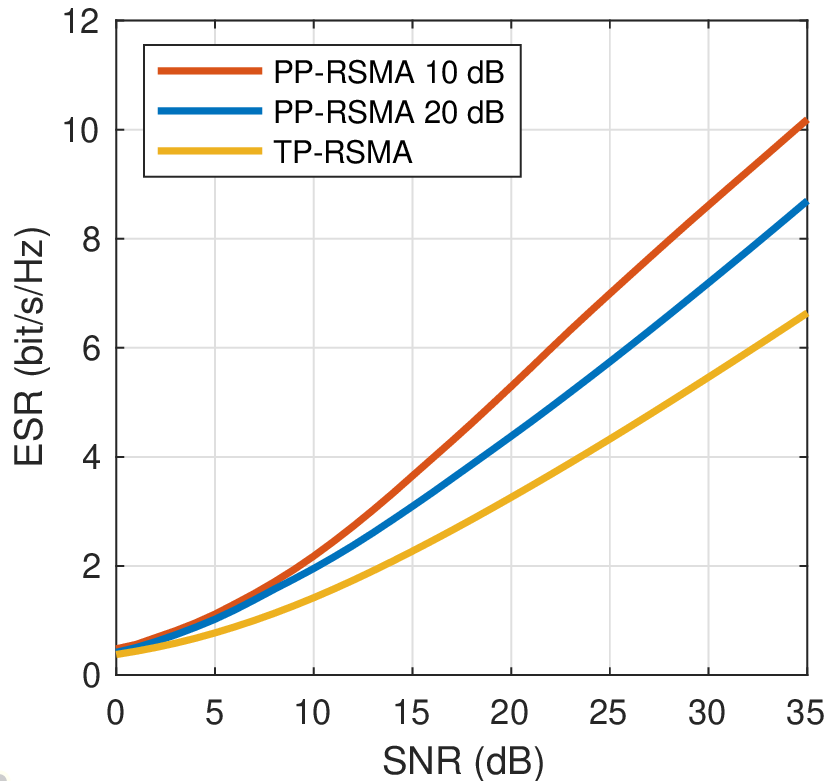}%
		\caption{Random and ZF precoding}
	\end{subfigure}%
	\caption{Ergodic sum rate of $\alpha$-users, $M=K_{\alpha}=2$, $K_0=1$, $\alpha=0.5$ and $\theta=0.5$.  Same rate for the $0$-user is maintained for the cases when the long-term SNR of the $0$-user is $10$ dB and $20$ dB lower than  that of user 1 and user 2.  }
	\label{fig:Plot_1 b}
\end{figure}
We first study the ESR versus SNR result for the two proposed approaches with optimized precoding in Fig. \ref{fig:Plot_1 b}(a)  and we compare it with random and ZF precoding as illustrated in Fig. \ref{fig:Plot_1 b}(b). The BS has $M=2$ transmit antennas and serves two single-antenna $\alpha$-users and one single-antenna $0$-users. $\alpha=\theta=0.5$. Same as in Fig. \ref{fig:Plot_1 a}, the ER of  $0$-user is maintained to be the same for both TP--RSMA and PP--RSMA and the long-term SNR of the $0$-user is 10 dB or 20 dB lower than that of the $\alpha$-users.  As the precoders are  optimally designed, the ESR of each approach is boosted in Fig. \ref{fig:Plot_1 b}(a) compared with the  result in Fig. \ref{fig:Plot_1 b}(b). The ESR performance when the precoders are optimized coincides with the performance when  random and ZF precoding is used.  The ESR gain of PP--RSMA over TP--RSMA still increases with the SNR difference between the two approaches.

\begin{figure}
	\centering
	\begin{subfigure}[b]{0.33\textwidth}
		\vspace{-1mm}
		\centering
		\includegraphics[width=0.88\textwidth]{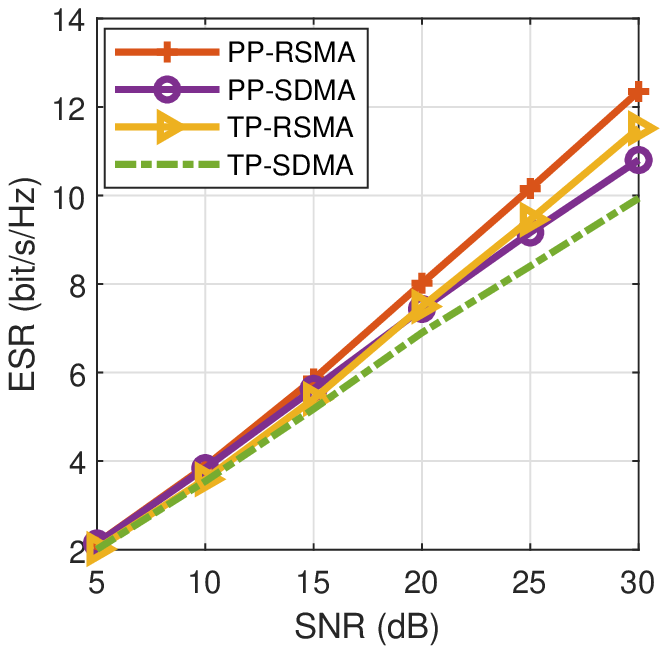}%
		\caption{$\mathbf{r}_0^{th}=[0.04,0.1,0.2,0.3,0.5,0.7]$, $R_{\alpha}^{th}=0$ bit/s/Hz  }
	\end{subfigure}%
	~
	\begin{subfigure}[b]{0.33\textwidth}
		\vspace{-1mm}
		\centering
		\includegraphics[width=0.88\textwidth]{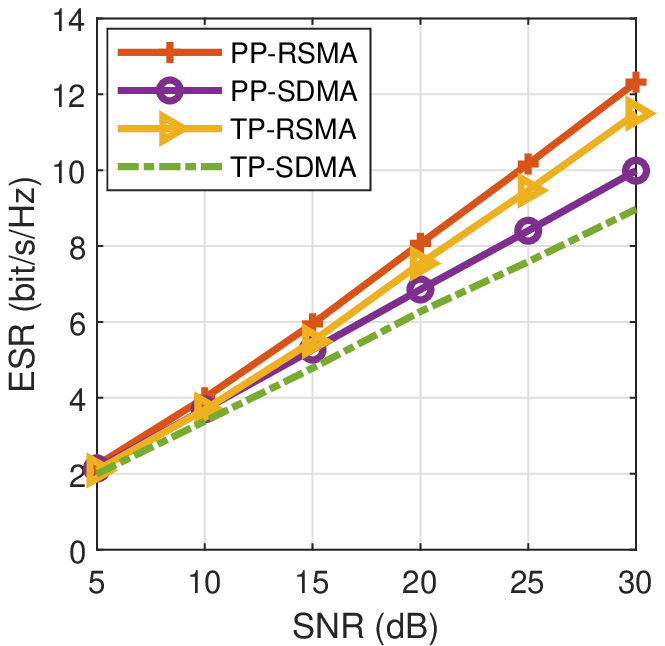}%
		\caption{$\mathbf{r}_0^{th}=[0.04,0.1,0.2,0.3,0.5,0.7]$, $\mathbf{r}_{\alpha}^{th}=[0.1,0.2 \cdots 0.6]$ bit/s/Hz }
	\end{subfigure}%
	~
	\begin{subfigure}[b]{0.32\textwidth}
		\vspace{-1mm}
		\centering
		\includegraphics[width=0.91\textwidth]{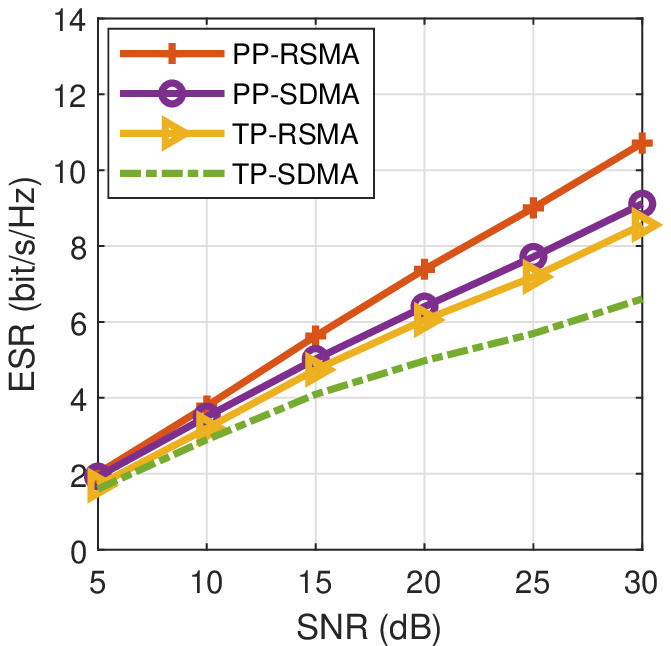}%
		\caption{$\mathbf{r}_0^{th}=[0.1,0.2,0.4,0.8,1.4,2]$, $\mathbf{r}_{\alpha}^{th}=[0.1,0.2 \cdots 0.6]$ bit/s/Hz}
	\end{subfigure}%
	
	\vspace{-1mm}
	\caption{Ergodic sum rate  of $\alpha$-users versus SNR comparison of different strategies, $M=2$, $K=3$, $\alpha=0.5$.}
	\label{fig: ESR vs SNR 1}
\end{figure}

Next, we take TP factor optimization into consideration and compare the four strategies illustrated in Fig. \ref{fig: schemeComparison}. We assume that there is  a 10 dB SNR difference between TP--RSMA and PP--RSMA approaches in all the following results. Without loss of generality, the QoS rate constraints of all $\alpha$-users are assumed to be equal, i.e., $R_{k}^{th}=R_{\alpha}^{th},\forall k\in\mathcal{K}_{\alpha}$ and that of all $0$-users are assumed to be equal, i.e., $R_{k}^{th}=R_{0}^{th},\forall k\in\mathcal{K}_{0}$.  We first study the influence of  QoS rate constraints when there are  two $\alpha$-users with CSIT quality factor $\alpha=0.5$ and one $0$-user.  As SNR increases from 5 to 30 dB, the QoS rate constraint for each $0$-user increases as $\mathbf{r}_0^{th}=[0.04,0.1,0.2,0.3,0.5,0.7]$ bit/s/Hz in Fig. \ref{fig: ESR vs SNR 1}(a)  and Fig. \ref{fig: ESR vs SNR 1}(b)  or increases as $\mathbf{r}_0^{th}=[0.1,0.2,0.4,0.8,1.4,2]$ bit/s/Hz in Fig. \ref{fig: ESR vs SNR 1}(c). The  QoS rate constraint for each $\alpha$-user is 0 for all considered SNR  in Fig. \ref{fig: ESR vs SNR 1}(a) while it is equal to $\mathbf{r}_{\alpha}^{th}=[0.1,0.2,0.3,0.4,0.5,0.6]$ bit/s/Hz for the corresponding SNR values in Fig. \ref{fig: ESR vs SNR 1}(b) and Fig. \ref{fig: ESR vs SNR 1}(c).
 In all subfigures of Fig. \ref{fig: ESR vs SNR 1}, PP--RSMA achieves non-negligible ESR gain over all other strategies.  
 Comparing Fig. \ref{fig: ESR vs SNR 1}(a) and Fig. \ref{fig: ESR vs SNR 1}(b), we observe that the ESR gain of RSMA-based strategies (including PP--RSMA and TP--RSMA) over SDMA-based strategies (including PP--SDMA and TP--SDMA)  increases with  the QoS rate constraint of  $\alpha$-users. The tendency coincides with the performance of $\alpha$-users-only transmission illustrated in \cite{RS2016hamdi}. By enabling the common stream for $\alpha$-users, RS is capable of managing interference more flexibly and therefore achieves non-negligible performance gain towards varied QoS rate constraints of the $\alpha$-users.  Comparing Fig. \ref{fig: ESR vs SNR 1}(b) and Fig. \ref{fig: ESR vs SNR 1}(c), we notice that the performance of TP-based strategies (TP--RSMA and TP--SDMA) drops dramatically with the QoS rate constraint of the $0$-users. In other words, PP-based strategies achieves higher ESR performance improvement as the QoS rate constraints of $0$-users increase. By embracing the benefits of PP-based and RSMA-based strategies, PP--RSMA achieves appealing ESR performance improvement over all other schemes towards  various QoS rate constraints of  $\alpha$-users and $0$-users.

\begin{figure}
	\centering
		\vspace{-4mm}
	\begin{subfigure}[b]{0.31\textwidth}
		\vspace{-1mm}
		\centering
		\includegraphics[width=0.91\textwidth]{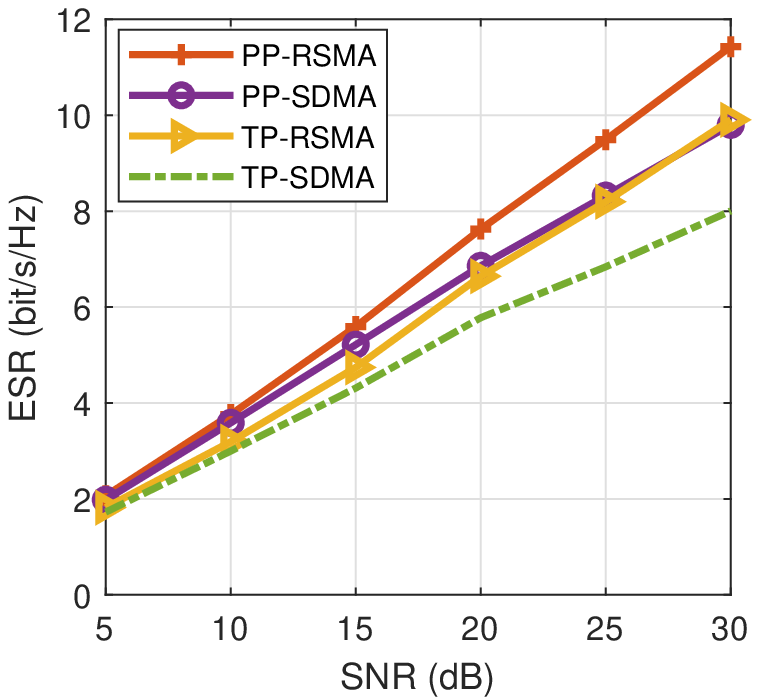}%
		\caption{$M=2$, $K=4$ }
	\end{subfigure}%
	~
	\begin{subfigure}[b]{0.31\textwidth}
		\vspace{-1mm}
		\centering
		\includegraphics[width=0.91\textwidth]{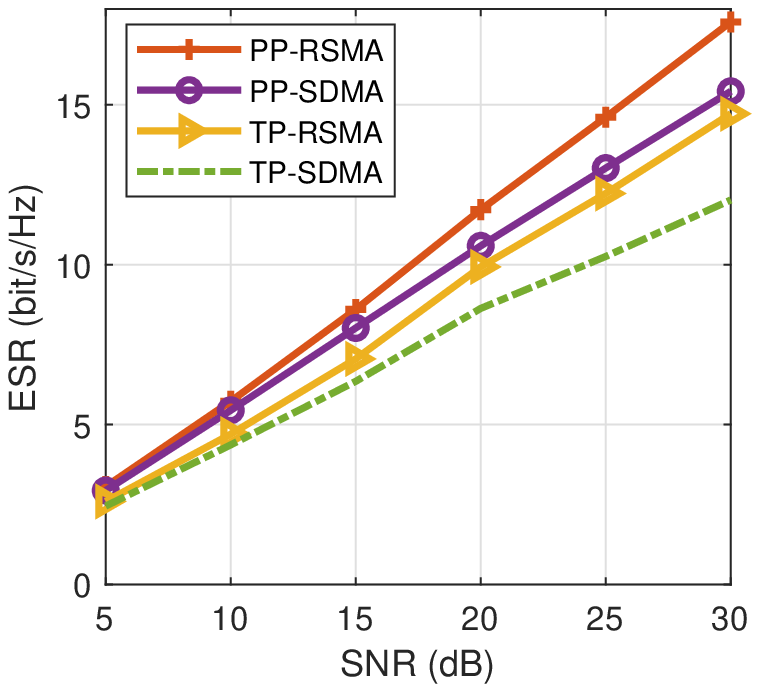}%
		\caption{$M=4$, $K=6$ }
	\end{subfigure}%
	~
	\begin{subfigure}[b]{0.31\textwidth}
		\vspace{-1mm}
		\centering
		\includegraphics[width=0.91\textwidth]{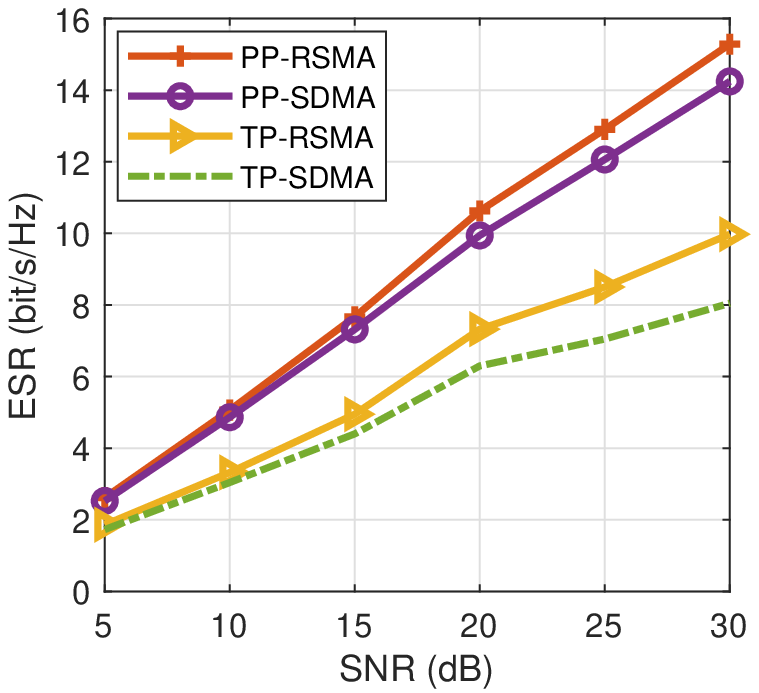}%
		\caption{$M=4$, $K=8$}
	\end{subfigure}%
	
	\vspace{-2mm}
	\caption{Ergodic sum rate of $\alpha$-users versus SNR comparison of different strategies,  $\alpha=0.5$, $\mathbf{r}_0^{th}=[0.04,0.1,0.2,0.3,0.5,0.7]$ bit/s/Hz, $R_{\alpha}^{th}=0.1$ bit/s/Hz. }
	\label{fig: ESR vs SNR 2}
\end{figure}
We further study the influence of different network loads  in Fig. \ref{fig: ESR vs SNR 2}. 
 Comparing Fig. \ref{fig: ESR vs SNR 2}(a) and Fig. \ref{fig: ESR vs SNR 2}(b),  the ESRs of all strategies increase with the number of $\alpha$-users. The ESR gaps between the corresponding PP-based strategies and TP-based strategies increase with the number of $\alpha$-users.  PP-based strategies are capable of serving more $\alpha$-users. Comparing Fig. \ref{fig: ESR vs SNR 2}(b) and Fig. \ref{fig: ESR vs SNR 2}(c), we also obtain that PP-based strategies are more suited to the cases with more $0$-users  and the ESR gap between PP--RSMA and PP--SDMA decreases with the number of $0$-users. As the number of $0$-users increases, a larger amount of power is allocated to $0$-users so as to meet their QoS rate constraints. Therefore, the remaining amount of power allocated to $\alpha$-users reduces. As the performance benefit of RSMA increases with SNR  \cite{mao2019TCOM}, adequate power allocation for the $\alpha$-streams allows RSMA to better determine the level of the interference to decode and treat as noise. 
Thanks to its ability of partially decoding interference,  partially treating interference as noise as well as non-orthogonal serving $\alpha$-and $0$-users, PP--RSMA achieves the highest ESR performance  over all other strategies even when the network load becomes much overloaded.

\begin{figure}
	\centering
	\begin{subfigure}[b]{0.4\textwidth}
		\vspace{-1mm}
		\centering
		\includegraphics[width=0.83\textwidth]{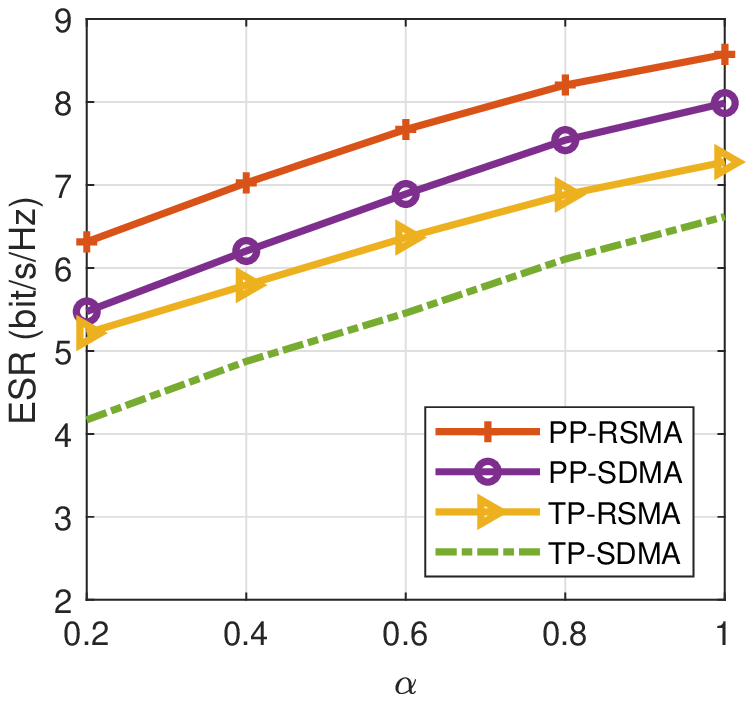}%
		\caption{ $R_{\alpha}^{th}=0.2$ bit/s/Hz }
	\end{subfigure}%
	~
	\begin{subfigure}[b]{0.4\textwidth}
		\vspace{-1mm}
		\centering
		\includegraphics[width=0.83\textwidth]{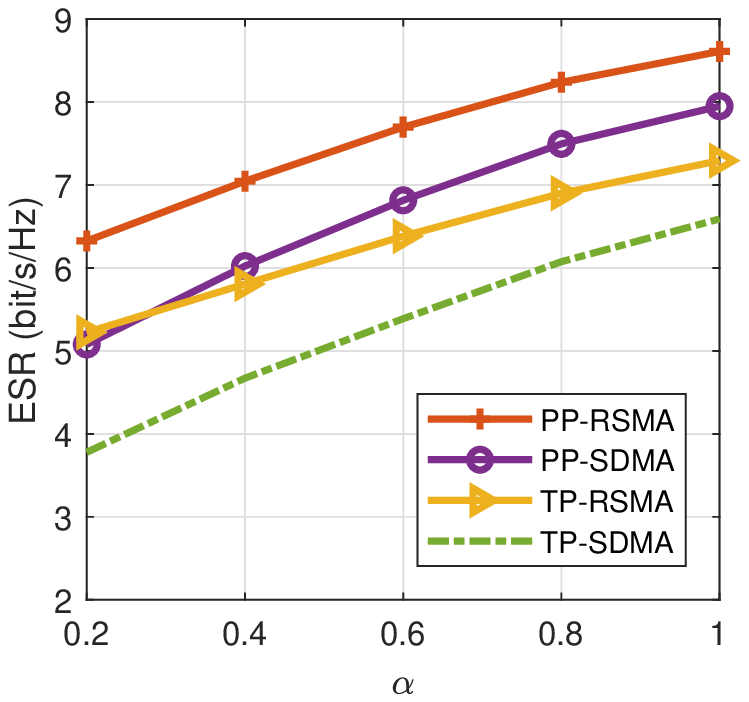}%
		\caption{$R_{\alpha}^{th}=0.4$ bit/s/Hz }
	\end{subfigure}%
	\caption{Ergodic sum rate of partial-CSIT  users versus the CSIT quality factor $\alpha$ comparison of different strategies, $M=2$,  $K=4$,  $\mathrm{SNR}=20 \,\mathrm{ dB}$, $R_0^{th}=0.4$ bit/s/Hz}
	\label{fig: ESR vs alpha}
\end{figure}
Fig. \ref{fig: ESR vs alpha} illustrates the ESR versus CSIT quality factor comparison of all the considered strategies. 
By comparing Fig. \ref{fig: ESR vs alpha}(a) and Fig. \ref{fig: ESR vs alpha}(b), we observe that as $\alpha$ decreases, the ESRs of SDMA-based strategies drop when the QoS rate constraints of $\alpha$-users increase while there is  no rate loss for the RSMA-based strategies. RSMA enhances user fairness and enables a more flexible allocation of user rates. 

All above, we conclude that PP--RSMA possesses the capability of both PP and RSMA, and becomes a more powerful and general transmission strategy for overloaded cellular IoT networks with heterogeneous CSIT quality and QoS rate constraints of all users. It is less sensitive to CSIT inaccuracy, and achieves a higher ESR gain over other strategies as the number of $\alpha$-and $0$-users increases or as the QoS rate constraint $R_k^{th}$ of each user  becomes larger.

\vspace{-0.2cm}
\section{Conclusion} \label{sec: conclusion}
To conclude, we consider an overloaded MISO BC with $M$ transmit antennas and $K$ users ($M<K$) where the transmitter has partial CSI for $M$ users  and statistical CSI for the remaining $K-M$ users. We incorporate  RS strategy into the considered cellular IoT-suited scenario and propose TP--RSMA and PP--RSMA approaches. PP--RSMA superposes the symbol for $0$-users on top of the symbols for $\alpha$-users. Such non-orthogonal transmission strategy has been shown to achieve strict DoF gains over TP--RSMA where the $0$-users are served independently from $\alpha$-users over orthogonal time slots. Moreover, the DoF region achieved by PP--RSMA has been shown to be optimal. Besides the DoF analysis in the high SNR regime, we also design the precoders of TP--RSMA and PP--RSMA by studying the ESR maximization problem subject to the QoS rate requirements of all users. Numerical results show that PP--RSMA achieves explicit sum rate  gain over TP--RSMA and all other baseline schemes. The DoF gain of PP--RSMA at high SNR is realized at finite SNR as well. 
Moreover, PP--RSMA is less sensitive to CSIT inaccuracy. The achieved ESR gap  over TP-based strategies increases with the number of $\alpha$-users and 0-users as well as QoS rate constraints of all users. PP--RSMA is suited to an extremely overloaded scenario.   According to the DoF and rate analysis, we draw the conclusion that PP--RSMA  is a powerful physical-layer transmission approach for overloaded cellular IoT.

 \vspace{-3mm}
\section*{Appendix I}
\section*{Proof of the optimum DoF region $\mathcal{D}$}
The DoF region $\mathcal{D}$ described by the inequalities in (\ref{dis_1}) and (\ref{dis_2}) is a $K$-dimensional polyhedron.
To prove the optimality of $\mathcal{D}$, we show that it is both achievable, and an outer bound of the optimal region.

\textit{Achievability}:
We first prove the achievability of $\mathcal{D}$.
Before we delve into the general case, we first characterize the achievable DoF region obtained by switching off all users in $\mathcal{K}_0$ (forcing their DoF to zero).
This is equivalent to projecting $\mathcal{D}$ onto the $M$ dimensional subspace characterized by $d_{M+1}, \cdots, d_K=0$.
It is readily seen that this setting corresponds to the $K$-User MISO
BC with partial CSIT in \cite{enrico2017bruno}, restricted to the case where all users have the same CSIT quality. 
The DoF region can be then obtained by \cite[Theorem 1]{enrico2017bruno} and it is  
given by the lemma below. 
This region is  used as a building block to prove the achievability of $\mathcal{D}$.
\begin{lemma} \label{prop_M=K}
	For a MISO BC with $K=M$ and CSIT quality $\alpha \in [0,1]$ for all users, an achievable DoF region $\mathcal{D}_{M=K}$ is given by
	\begin{equation} \label{dis_1_M=K}
	d_k \geq 0, \quad \forall k \in \mathcal{K}
	\end{equation}
	\begin{equation} \label{dis_2_M=K}
	\sum_{k \in \mathcal{S}}{d_{k}}  \leq 1 + (|\mathcal{S}|-1) \alpha, \quad \forall \mathcal{S} \subseteq  \mathcal{K}, |\mathcal{S}| \geq 1
	\end{equation}
	where $\mathcal{K}$ denotes the set of users $\{1,\ldots,K\}$.
\end{lemma}
We can now proceed to show the achievability of the region $\mathcal{D}$.
First, defining $d_{\Sigma}=\sum_{i \in \mathcal{K}_{0}}{d_i}$, the problem is equivalent to showing that all the
non-negative tuples $(d_1,\ldots,d_M,d_{\Sigma})$ that satisfy
\begin{equation} \label{dis_1_proof}
d_i \geq 0, d_{\Sigma} \geq 0 \quad \forall i \in \mathcal{K}_{\alpha}
\end{equation}
\begin{equation} \label{dis_2_proof}
\sum_{i \in \mathcal{S}}{d_{i}} +  d_{\Sigma}  \leq 1 + (|\mathcal{S}|-1) \alpha, \quad \forall \mathcal{S} \subseteq  \mathcal{K}_{\mathrm{\alpha}}, |\mathcal{S}| \geq 1
\end{equation}
are achievable. All tuples $(d_1,\ldots,d_M,d_{M+1},\cdots,d_K)$ are then obtained by splitting, in all
possible variants, the values of $d_{\Sigma}$ among users in $\mathcal{K}_0$.
The proof follows similar steps as in \cite{enrico2017bruno} but, in this case, the induction is done over the number of users in $\mathcal{K}_{\mathrm{\alpha}}$, denoted as $K_{\alpha}$ and equal to $M$. The case $K_{\alpha}=1$ is trivial. We assume that the hypothesis holds for
$K_{\alpha}=1,\ldots,k-1$.
 As in \cite{enrico2017bruno}, we show that each facet of the polyhedron is achievable.
Starting with the hyperplanes in (\ref{dis_2_proof}), for each subset $\mathcal{S} \subseteq \mathcal{K}_{\alpha}, |\mathcal{S}| \geq 1$,
we need to show that all the non-negative tuples $(d_1,\ldots,d_k,d_{\Sigma})$ that satisfy
\begin{equation*}
\begin{cases}
\sum_{i \in \mathcal{S}}{d_{i}} + d_{\Sigma} = 1 + (|\mathcal{S}|-1) \alpha\\
\sum_{i \in \bar{\mathcal{S}}}{d_{i}} + d_{\Sigma} \leq 1 + (|\bar{\mathcal{S}|}-1) \alpha, \forall {\bar{\mathcal{S}}} \subseteq {\mathcal{K}}_{\alpha}, \bar{\mathcal{S}} \neq \mathcal{S}, |\bar{\mathcal{S}}| \geq 1
\end{cases}
\end{equation*}
are achievable.
Following the same steps as in \cite{enrico2017bruno}, it can be verified that the above conditions are equivalent to
\begin{equation*}
\begin{cases}
\sum_{i \in \mathcal{S}}{d_{i}} + d_{\Sigma} = 1 + (|\mathcal{S}|-1) \alpha\\
d_i \geq \alpha, & \forall i \in \mathcal{S}\\
d_i \leq \alpha, & \forall i  \in \mathcal{K}_{\mathrm{\alpha}} \setminus \mathcal{S}.
\end{cases}
\end{equation*}
Each DoF tuple is achieved through power partitioning by allocating powers
scaling as $O(P^{\alpha})$ to private symbols of users $i \in \mathcal{S}$, and
powers scaling as $O(P^{d_{i}})$  to private symbols of users $i \in \mathcal{K}_{\mathrm{\alpha}} \setminus \mathcal{S}$.
On top, we consider all possible power partitions $\beta \in [\alpha,1]$ and for each partition, the common symbol's DoF is split, in all possible variants, among users $k \in \mathcal{S}$ only, while $d_{\Sigma}=1-\beta$.

Considering the facets contained in the hyperplanes in (\ref{dis_2_proof}),
we have two cases.
The first is given by $d_{\Sigma}=0$ and it reduces to $k$ users with CSIT $\alpha$
and $k$ antennas as in Lemma \ref{prop_M=K}. The second case considers any $j \in \mathcal{K}_{\alpha}$ and we have
\begin{equation*}
\begin{cases}
d_j=0\\
\sum_{i \in \mathcal{S}}{d_{i}} + d_{\Sigma}  \leq 1 + (|\mathcal{S}|-1) \alpha, \forall \mathcal{S}\subseteq \mathcal{K}_{\mathrm{\alpha}} \setminus \{j\}, |\mathcal{S}| \geq 1. \\
\end{cases}
\end{equation*}
This corresponds to the region in (\ref{dis_1_proof}) and (\ref{dis_2_proof}) considering the $k-1$ users in $\mathcal{K}_{\alpha}$. It is readily seen that this region is achievable by induction. As all facets of the polyhedron are achievable, all the remaining points can be achieved by time-sharing and the region is achievable.

\textit{Converse}:
The converse is based on the sum-DoF upperbound obtained in \cite{AG2015}.
For an arbitrary subset of users $\mathcal{U} \subseteq \mathcal{K}$, the sum-DoF is upperbounded by
\vspace{-1mm}
\begin{equation}
\label{eq_sum_DoF_UB}
\sum_{k \in \mathcal{U} }{d_k} \leq 1 + \alpha(|\mathcal{S}|-1)^+
\vspace{-1mm}
\end{equation}
where $\mathcal{S}=\mathcal{U} \cap \mathcal{K}_{\mathrm{\alpha}}$.
We increase the number of transmitter antennas to $K$ and then enhance the quality of one of the users in $\mathcal{S}$ to $1$ (if $\mathcal{S}$ is empty we pick any other user).
Since the previous steps provide an outerbound and cannot harm the DoF, \eqref{eq_sum_DoF_UB} directly follows from \cite[Theorem 1]{AG2015}.
By removing all redundant inequalities, the outerbound coincides with the region $\mathcal{D}$.


\vspace{-0.4cm}
\section*{Acknowledgement}
The authors are deeply indebted to Dr. Hamdi Joudeh for his helpful
comments and suggestions.

\vspace{-0.7cm}
\bibliographystyle{IEEEtran}
\bibliography{reference}

\end{document}